\documentclass[final,5p,sort&compress,times,letterpaper]{elsarticle}
\usepackage{amsmath,amssymb}
\usepackage{graphicx}
\usepackage[draft]{hyperref}
\usepackage{lipsum, color}
\usepackage{epsfig}
\usepackage{xspace}
\usepackage{dsfont}
\usepackage{widetext} 

\begin{document}
\newlength{\wdo}
\newcommand{\stroke}[1]{{$#1$}%
\settowidth{\wdo}{${#1}$} {\kern-\wdo}%
\partialvartstrokedint}

\makeatletter
\newcommand{\fancysep}{%
  \@afterindentfalse
  {\begin{center}
    \resizebox{0.8\linewidth}{0.4ex}{{%
        \fontsize{20}{24}\usefont{U}{webo}{xl}{n}{4}}}%
  \end{center}}\@afterheading}
\makeatother

\newcommand\reallywidetilde[1]{%
\begin{array}{c}
\stretchto{
  \scaleto{
    \scalerel*[\widthof{#1}]{\sim}
    {\rule[-\textheight/2]{1ex}{\textheight}} 
  }{1.7\textheight} 
}{0.6ex}\\           
\ensuremath{#1} \\                 
\rule{-1ex}{0ex}
\end{array}
}

\newenvironment{ibox}[1]%
{\vskip 1.0em
\framebox[\columnwidth][r]{%
\begin{minipage}[c]{\columnwidth}%
\vspace{-1.0em}%
#1%
\end{minipage}}}
{\vskip 1.0em}

\newcommand{\iboxed}[1]{%
\vskip 1.0em
\framebox[\columnwidth][r]{%
\begin{minipage}[c]{\columnwidth}%
\vspace{-1.0em}
#1%
\end{minipage}}
\vskip 1.0em}

\newcommand{\fitbox}[2]{%
\vskip 1.0em
\begin{flushright}
\framebox[{#1}][r]{%
\begin{minipage}[c]{\columnwidth}%
\vspace{-1.0em}
#2%
\end{minipage}}
\end{flushright}
\vskip 1.0em}

\newcommand{\iboxeds}[1]{%
\vskip 1.0em
\begin{equation}
\fbox{%
\begin{minipage}[c]{1mm}%
\vspace{-1.0em}
#1%
\end{minipage}}
\end{equation}
\vskip 1.0em}

\def\Xint#1{\mathchoice
   {\XXint\displaystyle\textstyle{#1}}%
   {\XXint\textstyle\scriptstyle{#1}}%
   {\XXint\scriptstyle\scriptscriptstyle{#1}}%
   {\XXint\scriptscriptstyle\scriptscriptstyle{#1}}%
   \!\int}
\def\XXint#1#2#3{{\setbox0=\hbox{$#1{#2#3}{\int}$}
     \vcenter{\hbox{$#2#3$}}\kern-.5\wd0}}
\def\ddashint{\Xint=}
\def\dashint{\Xint-}

\def\mathbi#1{\textbf{\em #1}}

\newcommand{\alps}{\ensuremath{\alpha_s}}
\newcommand{\qbar}{\bar{q}}
\newcommand{\ubar}{\bar{u}}
\newcommand{\dbar}{\bar{d}}
\newcommand{\sbar}{\bar{s}}
\newcommand{\Sbar}{\overline{S}}
\newcommand{\beq}{\begin{equation}}
\newcommand{\eeq}{\end{equation}}
\newcommand{\beqa}{\begin{eqnarray}}
\newcommand{\eeqa}{\end{eqnarray}}
\newcommand{\gs}{g_{\pi NN}}
\newcommand{\gw}{f_\pi}
\newcommand{\mq}{m_Q}
\newcommand{\mn}{m_N}
\newcommand{\mpi}{m_\pi}
\newcommand{\mrho}{m_\rho}
\newcommand{\momg}{m_\omega}
\newcommand{\bb}{\langle}
\newcommand{\kb}{\rangle}
\newcommand{\xvec}{\mathbf{x}}
\newcommand{\st}{\ensuremath{\sqrt{\sigma}}}
\newcommand{\Bvec}{\mathbf{B}}
\newcommand{\dext}{\mathbf{d}}
\newcommand{\rvec}{\mathbf{r}}
\newcommand{\Rvec}{\mathbf{R}}
\newcommand{\kvec}{\mathbf{k}}
\newcommand{\pvec}{\mathbf{p}}
\newcommand{\qvec}{\mathbf{q}}
\newcommand{\Pvec}{\mathbf{P}}
\newcommand{\vvec}{\mathbf{v}}
\newcommand{\Vvec}{\mathbf{V}}
\newcommand{\Fvec}{\mathbf{F}}
\newcommand{\bvec}[1]{\ensuremath{\mathbf{#1}}}
\newcommand{\bvh}[1]{\ensuremath{\hat{\mathbf{#1}}}}
\newcommand{\bra}[1]{\ensuremath{\bb#1|}}
\newcommand{\ket}[1]{\ensuremath{|#1\kb}}
\newcommand{\pbra}[1]{\ensuremath{(#1|}}
\newcommand{\pket}[1]{\ensuremath{|#1)}}
\newcommand{\gft}{\ensuremath{\gamma_{FT}}}
\newcommand{\gfv}{\ensuremath{\gamma_5}}
\newcommand{\bfalp}{\ensuremath{\bm{\alpha}}}
\newcommand{\bfbeta}{\ensuremath{\bm{\beta}}}
\newcommand{\bfeps}{\ensuremath{\bm{\epsilon}}}
\newcommand{\lag}{{\lambda_\gamma}}
\newcommand{\lao}{{\lambda_\omega}}
\newcommand{\lN}{\lambda_N}
\newcommand{\lM}{\lambda_M}
\newcommand{\lB}{\lambda_B}
\newcommand{\epslag}{\ensuremath{\bm{\epsilon}_{\lag}}}
\newcommand{\bfept}{\ensuremath{\tilde{\bm{\epsilon}}}}
\newcommand{\bfgam}{\ensuremath{\bm{\gamma}}}
\newcommand{\bfnab}{\ensuremath{\bm{\nabla}}}
\newcommand{\bflambda}{\ensuremath{\bm{\lambda}}}
\newcommand{\bfmu}{\ensuremath{\bm{\mu}}}
\newcommand{\bfphi}{\ensuremath{\bm{\phi}}}
\newcommand{\bfvphi}{\ensuremath{\bm{\varphi}}}
\newcommand{\bfpi}{\ensuremath{\bm{\pi}}}
\newcommand{\bfsig}{\ensuremath{\bm{\sigma}}}
\newcommand{\bftau}{\ensuremath{\bm{\tau}}}
\newcommand{\bfpsi}{\ensuremath{\bm{\psi}}}
\newcommand{\bfdelta}{\ensuremath{\bm{\delta}}}
\newcommand{\bfrho}{\ensuremath{\bm{\rho}}}
\newcommand{\bfth}{\ensuremath{\bm{\theta}}}
\newcommand{\bfchi}{\ensuremath{\bm{\chi}}}
\newcommand{\bfxi}{\ensuremath{\bm{\xi}}}
\newcommand{\bfR}{\ensuremath{\bvec{R}}}
\newcommand{\bfP}{\ensuremath{\bvec{P}}}
\newcommand{\bfJ}{{\mathbi{J}}}
\newcommand{\bfj}{{\mathbi{j}}}
\newcommand{\Rcm}{\ensuremath{\bvec{R}_{CM}}}
\newcommand{\spup}{\uparrow}
\newcommand{\spd}{\downarrow}
\newcommand{\up}{\uparrow}
\newcommand{\dn}{\downarrow}
\newcommand{\hbarom}{\frac{\hbar^2}{m_Q}}
\newcommand{\half}{\ensuremath{\frac{1}{2}}}
\newcommand{\thalf}{\ensuremath{\frac{3}{2}}}
\newcommand{\fhalf}{\ensuremath{\frac{5}{2}}}
\newcommand{\shalf}{\ensuremath{{\tfrac{1}{2}}}}
\newcommand{\sqtr}{\ensuremath{{\tfrac{1}{4}}}}
\newcommand{\sphalf}{\ensuremath{\genfrac{}{}{0pt}{1}{+}{}\!\tfrac{1}{2}}}
\newcommand{\smhalf}{\ensuremath{\genfrac{}{}{0pt}{1}{-}{}\!\tfrac{1}{2}}}
\newcommand{\sthalf}{\ensuremath{{\tfrac{3}{2}}}}
\newcommand{\spthalf}{\ensuremath{{\tfrac{+3}{2}}}}
\newcommand{\smthalf}{\ensuremath{{\tfrac{-3}{2}}}}
\newcommand{\sfhalf}{\ensuremath{\tfrac{5}{2}}}
\newcommand{\sshalf}{\ensuremath{\tfrac{7}{2}}}
\newcommand{\mythird}{\ensuremath{\frac{1}{3}}}
\newcommand{\tthird}{\ensuremath{\frac{2}{3}}}
\newcommand{\sthird}{\ensuremath{\tfrac{1}{3}}}
\newcommand{\stthird}{\ensuremath{\tfrac{2}{3}}}
\newcommand{\vnn}{\ensuremath{\hat{v}_{NN}}}
\newcommand{\vij}{\ensuremath{\hat{v}_{ij}}}
\newcommand{\vik}{\ensuremath{\hat{v}_{ik}}}
\newcommand{\argonne}{\ensuremath{v_{18}}}
\newcommand{\lqcd}{\ensuremath{\mathcal{L}_{QCD}}}
\newcommand{\lqed}{\ensuremath{\mathscr{L}_{QED}}}
\newcommand{\Lttp}{\ensuremath{\mathscr{L}_{tt'}}\xspace}
\newcommand{\Lttf}{\ensuremath{\mathscr{L}_{tt',0}}\xspace}
\newcommand{\Lttint}{\ensuremath{\mathscr{L}_{tt',\mbox{\scriptsize int}}}\xspace}
\newcommand{\lgf}{\ensuremath{\mathcal{L}_g}}
\newcommand{\lqm}{\ensuremath{\mathcal{L}_q}}
\newcommand{\lqg}{\ensuremath{\mathcal{L}_{qg}}}
\newcommand{\nn}{\ensuremath{N\!N}}
\newcommand{\nnn}{\ensuremath{N\!N\!N}}
\newcommand{\qq}{\ensuremath{qq}}
\newcommand{\qqq}{\ensuremath{qqq}}
\newcommand{\qqb}{\ensuremath{q\bar{q}}}
\newcommand{\hpnd}{\ensuremath{H_{\pi N\Delta}}}
\newcommand{\hpqq}{\ensuremath{H_{\pi qq}}}
\newcommand{\hpqqa}{\ensuremath{H^{(a)}_{\pi qq}}}
\newcommand{\hpqqe}{\ensuremath{H^{(e)}_{\pi qq}}}
\newcommand{\hint}{\ensuremath{H_{\rm int}}}
\newcommand{\fpnn}{\ensuremath{f_{\pi\! N\!N}}}
\newcommand{\fenn}{\ensuremath{f_{\eta\! N\!N}}}
\newcommand{\gsnn}{\ensuremath{g_{\sigma\! N\!N}}}
\newcommand{\gpnn}{\ensuremath{g_{\pi\! N\!N}}}
\newcommand{\fpnd}{\ensuremath{f_{\pi\! N\!\Delta}}}
\newcommand{\grpg}{\ensuremath{g_{\rho\pi\gamma}}}
\newcommand{\gopg}{\ensuremath{g_{\omega\pi\gamma}}}
\newcommand{\fmqq}{\ensuremath{f_{M\! qq}}}
\newcommand{\gmqq}{\ensuremath{g_{M\! qq}}}
\newcommand{\fpqq}{\ensuremath{f_{\pi qq}}}
\newcommand{\gpqq}{\ensuremath{g_{\pi qq}}}
\newcommand{\feqq}{\ensuremath{f_{\eta qq}}}
\newcommand{\gonn}{\ensuremath{g_{\omega N\!N}}}
\newcommand{\gonna}{\ensuremath{g^t_{\omega N\!N}}}
\newcommand{\grnn}{\ensuremath{g_{\rho N\!N}}}
\newcommand{\gopr}{\ensuremath{g_{\omega\pi\rho}}}
\newcommand{\grnp}{\ensuremath{g_{\rho N\!\pi}}}
\newcommand{\grpp}{\ensuremath{g_{\rho\pi\pi}}}
\newcommand{\gt}{\ensuremath{g_t}\xspace}
\newcommand{\gtp}{\ensuremath{g_{t'}}\xspace}
\newcommand{\Lpnn}{\ensuremath{\Lambda_{\pi\! N\! N}}}
\newcommand{\Lonn}{\ensuremath{\Lambda_{\omega N\! N}}}
\newcommand{\Lonna}{\ensuremath{\Lambda^t_{\omega N\! N}}}
\newcommand{\Lrnn}{\ensuremath{\Lambda_{\rho N\! N}}}
\newcommand{\Lopr}{\ensuremath{\Lambda_{\omega\pi\rho}}}
\newcommand{\Lrpp}{\ensuremath{\Lambda_{\rho\pi\pi}}}
\newcommand{\getaqq}{\ensuremath{g_{\eta qq}}}
\newcommand{\fsqq}{\ensuremath{f_{\sigma qq}}}
\newcommand{\gsqq}{\ensuremath{g_{\sigma qq}}}
\newcommand{\piqq}{\ensuremath{{\pi\! qq}}}
\newcommand{\ylm}{\ensuremath{Y_\ell^m}}
\newcommand{\ylmc}{\ensuremath{Y_\ell^{m*}}}
\newcommand{\ebh}[1]{\hat{\bvec{e}}_{#1}}
\newcommand{\kbh}{\hat{\bvec{k}}}
\newcommand{\nbh}{\hat{\bvec{n}}}
\newcommand{\pvbh}{\hat{\bvec{p}}}
\newcommand{\qbh}{\hat{\bvec{q}}}
\newcommand{\Xbh}{\hat{\bvec{X}}}
\newcommand{\rbh}{\hat{\bvec{r}}}
\newcommand{\xbh}{\hat{\bvec{x}}}
\newcommand{\ybh}{\hat{\bvec{y}}}
\newcommand{\zbh}{\hat{\bvec{z}}}
\newcommand{\betabh}{\hat{\bfbeta}}
\newcommand{\tbh}{\hat{\bfth}}
\newcommand{\pbh}{\hat{\bfvphi}}
\newcommand{\dt}{\Delta\tau}
\newcommand{\kmag}{|\bvec{k}|}
\newcommand{\pmag}{|\bvec{p}|}
\newcommand{\qmag}{|\bvec{q}|}
\newcommand{\oas}{\ensuremath{\mathcal{O}(\alpha_s)}}
\newcommand{\vtxb}{\ensuremath{\Lambda_\mu(p',p)}}
\newcommand{\vtxp}{\ensuremath{\Lambda^\mu(p',p)}}
\newcommand{\pwqp}{e^{i\bvec{q}\cdot\bvec{r}}}
\newcommand{\pwqm}{e^{-i\bvec{q}\cdot\bvec{r}}}
\newcommand{\gsa}[1]{\ensuremath{\bb#1\kb_0}}
\newcommand{\oer}[1]{\mathcal{O}\left(\frac{1}{\qmag^{#1}}\right)}
\newcommand{\nub}[1]{\overline{\nu^{#1}}}
\newcommand{\epf}{E_\bvec{p}}
\newcommand{\epfp}{E_{\bvec{p}'}}
\newcommand{\eka}{E_{\alpha\kappa}}
\newcommand{\ekaq}{(E_{\alpha\kappa})^2}
\newcommand{\ekap}{E_{\alpha'\kappa}}
\newcommand{\ekpa}{E+{\alpha\kappa_+}}
\newcommand{\ekma}{E_{\alpha\kappa_-}}
\newcommand{\ekp}{E_{\kappa_+}}
\newcommand{\ekm}{E_{\kappa_-}}
\newcommand{\ekpap}{E_{\alpha'\kappa_+}}
\newcommand{\ekmap}{E_{\alpha'\kappa_-}}
\newcommand{\yjm}[1]{\mathcal{Y}_{jm}^{#1}}
\newcommand{\ysa}[3]{\mathcal{Y}_{#1,#2}^{#3}}
\newcommand{\yjsl}[2]{\mathcal{Y}_{#1}^{#2}}
\newcommand{\yss}[2]{\mathcal{Y}_{#1}^{#2}}
\newcommand{\Dj}{\ensuremath{\mathscr{D}}}
\newcommand{\ysc}{\tilde{y}}
\newcommand{\enm}{\varepsilon_{NM}}
\newcommand{\Scg}[6]
	{\ensuremath{S^{#1}_{#4}\:\vphantom{S}^{#2}_{#5}
 	 \:\vphantom{S}^{#3}_{#6}\,}}
\newcommand{\Kmat}[6]
	{\ensuremath{K\left[\begin{array}{ccc} 
	#1 & #2 & #3 \\ #4 & #5 & #6\end{array}\right]}}
\newcommand{\irt}{\ensuremath{\frac{1}{\sqrt{2}}}}
\newcommand{\sirt}{\ensuremath{\tfrac{1}{\sqrt{2}}}}
\newcommand{\irth}{\ensuremath{\frac{1}{\sqrt{3}}}}
\newcommand{\sirth}{\ensuremath{\tfrac{1}{\sqrt{3}}}}
\newcommand{\irs}{\ensuremath{\frac{1}{\sqrt{6}}}}
\newcommand{\sirs}{\ensuremath{\tfrac{1}{\sqrt{6}}}}
\newcommand{\tors}{\ensuremath{\frac{2}{\sqrt{6}}}}
\newcommand{\stors}{\ensuremath{\tfrac{2}{\sqrt{6}}}}
\newcommand{\rtoth}{\ensuremath{\sqrt{\frac{2}{3}}}}
\newcommand{\rthot}{\ensuremath{\frac{\sqrt{3}}{2}}}
\newcommand{\ithrt}{\ensuremath{\frac{1}{3\sqrt{2}}}}
\newcommand{\Tg}{\ensuremath{\mathsf{T}}}
\newcommand{\irrep}[1]{\ensuremath{\mathbf{#1}}}
\newcommand{\cirrep}[1]{\ensuremath{\overline{\mathbf{#1}}}}
\newcommand{\Fij}{\ensuremath{\hat{F}_{ij}}}
\newcommand{\Fqij}{\ensuremath{\hat{F}^{(qq)}_{ij}}}
\newcommand{\Fsij}{\ensuremath{\hat{F}^{(qs)}_{ij}}}
\newcommand{\Opij}{\mathcal{O}^p_{ij}}
\newcommand{\fpij}{f_p(r_{ij})}
\newcommand{\titj}{\bftau_i\cdot\bftau_j}
\newcommand{\sisj}{\bfsig_i\cdot\bfsig_j}
\newcommand{\Sij}{S_{ij}}
\newcommand{\LS}{\bvec{L}_{ij}\cdot\bvec{S}_{ij}}
\newcommand{\TT}{\Tg_i\cdot\Tg_j}
\newcommand{\chet}{\ensuremath{\chi ET}}
\newcommand{\chpt}{\ensuremath{\chi PT}}
\newcommand{\chsy}{\ensuremath{\chi\mbox{symm}}}
\newcommand{\lchi}{\ensuremath{\Lambda_\chi}}
\newcommand{\lcon}{\ensuremath{\Lambda_{QCD}}}
\newcommand{\dcpsi}{\ensuremath{\bar{\psi}}}
\newcommand{\dcbfpsi}{\ensuremath{\bar{\bfpsi}}}
\newcommand{\dc}[1]{\ensuremath{\overline{#1}}}
\newcommand{\dcpsip}{\ensuremath{\bar{\psi}^{(+)}}}
\newcommand{\psip}{\ensuremath{{\psi}^{(+)}}}
\newcommand{\dcpsim}{\ensuremath{\bar{\psi}^{(-)}}}
\newcommand{\psim}{\ensuremath{{\psi}^{(-)}}}
\newcommand{\psit}{\ensuremath{\psi_t}}
\newcommand{\psitp}{\ensuremath{\psi_{t'}}}
\newcommand{\llo}{\ensuremath{\mathcal{L}^{(0)}_{\chet}}}
\newcommand{\lchet}{\ensuremath{\mathcal{L}_{\chi}}}
\newcommand{\hchet}{\ensuremath{\mathcal{H}_{\chi}}}
\newcommand{\Hd}{\ensuremath{\mathcal{H}}}
\newcommand{\Dmu}{\ensuremath{\mathcal{D}_\mu}}
\newcommand{\Dsl}{\ensuremath{\slashed{\mathcal{D}}}}
\newcommand{\comm}[2]{\ensuremath{\left[#1,#2\right]}}
\newcommand{\acomm}[2]{\ensuremath{\left\{#1,#2\right\}}}
\newcommand{\ev}[1]{\ensuremath{\bb\hat{#1}\kb}}
\newcommand{\exv}[1]{\ensuremath{\bb{#1}\kb}}
\newcommand{\evt}[1]{\ensuremath{\bb{#1}(\tau)\kb}}
\newcommand{\evm}[1]{\ensuremath{\bb{#1}\kb_M}}
\newcommand{\evv}[1]{\ensuremath{\bb{#1}\kb_V}}
\newcommand{\ovl}[2]{\ensuremath{\bb{#1}|{#2}\kb}}
\newcommand{\movl}[2]{\ensuremath{({#1}|{#2}\kb}}
\newcommand{\dmovl}[2]{\ensuremath{\bb{#1}|{#2})}}
\newcommand{\povl}[2]{\ensuremath{({#1}|{#2})}}
\newcommand{\tpfod}{\ovl{q_2 t_2}{q_1 t_1}\xspace}
\newcommand{\tpffi}{\ovl{q_f t_f}{q_i t_i}\xspace}
\newcommand{\tpf}[2]{\ovl{q_{#1} t_{#1}}{q_{#2} t_{#2}}\xspace}
\newcommand{\pd}{\partial}
\newcommand{\dndd}[2]{\ensuremath{\frac{d{{#1}}}{d{#2}}}}
\newcommand{\dpdd}[1]{\ensuremath{\frac{d\hphantom{{#1}}}{d{#1}}}}
\newcommand{\pnpd}[2]{\frac{\partial{#1}}{\partial{#2}}}
\newcommand{\pppd}[1]{\frac{\partial{\hphantom{#1}}}{\partial{#1}}}
\newcommand{\fnfd}[2]{\frac{\delta{#1}}{\delta{#2}}}
\newcommand{\rfnfd}[2]{\frac{\bfdelta{#1}}{\bfdelta{#2}}}
\newcommand{\fdfd}[1]{\frac{\delta}{\delta{#1}}}
\newcommand{\rfdfd}[1]{\frac{\overleftarrow{\delta}}{\delta{#1}}}
\newcommand{\brfdfd}[1]{\frac{{\bfdelta}}{\bfdelta{#1}}}
\newcommand{\plmu}{\partial_\mu}
\newcommand{\plnu}{\partial_\nu}
\newcommand{\pumu}{\partial^\mu}
\newcommand{\punu}{\partial^\nu}
\newcommand{\mcdf}{\delta^{(4)}(p_f-p_i-q)}
\newcommand{\ecdf}{\delta(E_f-E_i-\nu)}
\newcommand{\trace}{\mbox{Tr }}
\newcommand{\lxr}{\ensuremath{SU(2)_L\times SU(2)_R}}
\newcommand{\gV}[2]{\ensuremath{(\gamma^{-1})^{#1}_{\hphantom{#1}{#2}}}}
\newcommand{\gVd}[2]{\ensuremath{\gamma^{#1}_{\hphantom{#1}{#2}}}}
\newcommand{\LpV}[1]{\ensuremath{\Lambda^{#1}V}}
\newcommand{\hatH}{\ensuremath{\hat{H}}}
\newcommand{\hath}{\ensuremath{\hat{h}}}
\newcommand{\eht}{\ensuremath{e^{-\tau\hat{H}}}}
\newcommand{\ehdt}{\ensuremath{e^{-\Delta\tau\hat{H}}}}
\newcommand{\ehtm}{\ensuremath{e^{-\tau(\hat{H}-E_V)}}}
\newcommand{\ehdtm}{\ensuremath{e^{-\Delta\tau(\hat{H}-E_V)}}}
\newcommand{\Oop}{\ensuremath{\mathcal{O}}}
\newcommand{\Sop}{\ensuremath{\mathcal{S}}}
\newcommand{\Jop}{\ensuremath{\mathcal{J}}}
\newcommand{\Gop}{\ensuremath{\hat{\mathcal{G}}}}
\newcommand{\SU}[1]{\ensuremath{SU({#1})}}
\newcommand{\Un}[1]{\ensuremath{U({#1})}}
\newcommand{\proj}[1]{\ensuremath{\ket{#1}\bra{#1}}}
\newcommand{\su}[1]{\ensuremath{\mathfrak{su}({#1})}}
\newcommand{\ip}[2]{\ensuremath{\bvec{#1}\cdot\bvec{#2}}}
\newcommand{\norm}[1]{\ensuremath{\left| #1\right|^2}}
\newcommand{\rnorm}[1]{\ensuremath{\lvert #1\rvert}}
\newcommand{\pid}{\left(\begin{array}{cc} 1 & 0 \\ 0 & 1\end{array}\right)}
\newcommand{\psx}{\left(\begin{array}{cc} 0 & 1 \\ 1 & 0\end{array}\right)}
\newcommand{\psy}{\left(\begin{array}{cc} 0 & -i \\ i & 0\end{array}\right)}
\newcommand{\psz}{\left(\begin{array}{cc} 1 & 0 \\ 0 & -1\end{array}\right)}
\newcommand{\ua}{\uparrow}
\newcommand{\da}{\downarrow}
\newcommand{\deln}{\delta_{i_1 i_2\ldots i_n}}
\newcommand{\GabRR}{G_{\alpha\beta}(\bfR,\bfR')}
\newcommand{\GRR}{G(\bfR,\bfR')}
\newcommand{\GfRR}{G_0(\bfR,\bfR')}
\newcommand{\GRiR}{G(\bfR_i,\bfR_{i-1})}
\newcommand{\GRRs}[2]{G(\bfR_{#1},\bfR_{#2})}
\newcommand{\Gdgn}{\Gamma_{\Delta,\gamma N}}
\newcommand{\Gdgnb}{\overline\Gamma_{\Delta,\gamma N}}
\newcommand{\GJT}{\Gamma_{LS}^{JT}(k)}
\newcommand{\GJTa}[2]{\Gamma^{#1}_{#2}}
\newcommand{\GtwJTa}[2]{\tilde{\Gamma}_{#1}^{#2}}
\newcommand{\Gtw}{\tilde{\Gamma}}
\newcommand{\Gbar}{\overline{\Gamma}}
\newcommand{\Gtil}{\tilde{\Gamma}}
\newcommand{\Gpndb}{\overline{\Gamma}_{\pi N,\Delta}}
\newcommand{\GbNgn}{{\overline{\Gamma}}_{N^*,\gamma N}}
\newcommand{\GNgn}{\Gamma_{N^*,\gamma N}}
\newcommand{\GbNmb}{{\overline{\Gamma}}_{N^*,MB}}
\newcommand{\Lg}[2]{\ensuremath{L^{#1}_{\hphantom{#1}{#2}}}}
\newcommand{\psik}{\ensuremath{\left(\begin{matrix}\psi_1 \\ \psi_2\end{matrix}\right)}}
\newcommand{\psib}{\ensuremath{\left(\begin{matrix}\psi^*_1&\psi^*_2\end{matrix}\right)}}
\newcommand{\Gf}{\ensuremath{\frac{1}{E-H_0}}}
\newcommand{\Gv}{\ensuremath{\frac{1}{E-H_0-\vnres}}}
\newcommand{\Gx}{\ensuremath{\frac{1}{E-H_0-V}}}
\newcommand{\Gex}{\ensuremath{\mathcal{G}}}
\newcommand{\Gfpm}{\ensuremath{\frac{1}{E-H_0\pm i\epsilon}}}
\newcommand{\vres}{v_R}
\newcommand{\vnres}{v}
\newcommand{\tpz}{\ensuremath{^3P_0}}
\newcommand{\tres}{t_R}
\newcommand{\tsr}{t^R}
\newcommand{\tsnr}{t^{NR}}
\newcommand{\trest}{\tilde{t}_R}
\newcommand{\tnres}{t}
\newcommand{\Pt}{P_{12}}
\newcommand{\Sz}{\ket{S_0}}
\newcommand{\Sa}{\ket{S^{(-1)}_1}}
\newcommand{\Sb}{\ket{S^{(0)}_1}}
\newcommand{\Sc}{\ket{S^{(+1)}_1}}
\newcommand{\sbasis}{\ket{s_1 s_2; m_1 m_2}}
\newcommand{\Sbasis}{\ket{s_1 s_2; S M}}
\newcommand{\sket}[2]{\ket{{#1}\,{#2}}}
\newcommand{\sbra}[2]{\bra{{#1}\,{#2}}}
\newcommand{\psmket}{\ket{\bvec{p};s\,m}}
\newcommand{\cket}{\ket{\bvec{p};s_1 s_2\,m_1 m_2}}
\newcommand{\hket}{\ket{\bvec{p};s_1 s_2\,\lambda_1\lambda_2}}
\newcommand{\hkets}{\ket{s\,\lambda}}
\newcommand{\phkets}{\ket{\bvec{p};s\,\lambda}}
\newcommand{\klsjm}{\ket{p;\ell s; j m}}
\newcommand{\pq}{\bvec{p}_q}
\newcommand{\pqb}{\bvec{p}_{\qbar}}
\newcommand{\mps}[1]{\frac{d^3{#1}}{(2\pi)^{3/2}}}
\newcommand{\mpsf}[1]{\frac{d^3{#1}}{(2\pi)^{3}}}
\newcommand{\mpsfe}[1]{\frac{d^3p_{#1}}{2E_{#1}(2\pi)^{3}}}
\newcommand{\du}[1]{u_{\bvec{#1},s}}
\newcommand{\dv}[1]{v_{\bvec{#1},s}}
\newcommand{\cdu}[1]{\overline{u}_{\bvec{#1},s}}
\newcommand{\cdv}[1]{\overline{v}_{\bvec{#1},s}}
\newcommand{\dus}[2]{u_{\bvec{#1},{#2}}}
\newcommand{\dvs}[2]{v_{\bvec{#1},{#2}}}
\newcommand{\cdus}[2]{\overline{u}_{\bvec{#1},{#2}}}
\newcommand{\cdvs}[2]{\overline{v}_{\bvec{#1},{#2}}}
\newcommand{\bop}[1]{b_{\bvec{#1},s}}
\newcommand{\dop}[1]{d_{\bvec{#1},s}}
\newcommand{\bops}[2]{b_{\bvec{#1},{#2}}}
\newcommand{\dops}[2]{d_{\bvec{#1},{#2}}}
\newcommand{\mev}{\mbox{ MeV}}
\newcommand{\gev}{\mbox{ GeV}}
\newcommand{\fmi}{\mbox{ fm}}
\newcommand{\calM}{\mathcal{M}}
\newcommand{\Smat}{\mathcal{S}}
\newcommand{\JLSTh}{JLST\lambda}
\newcommand{\Tpg}{T_{\pi N,\gamma N}}
\newcommand{\tpg}{t_{\pi N,\gamma N}}
\newcommand{\vmbmb}{\ensuremath{v_{M'B',MB}}}
\newcommand{\tmbgn}{\ensuremath{t_{MB,\gamma N}}}
\newcommand{\Tonon}{\ensuremath{T_{\omega N,\omega N}}}
\newcommand{\tonon}{\ensuremath{t_{\omega N,\omega N}}}
\newcommand{\tronon}{\ensuremath{t^R_{\omega N,\omega N}}}
\newcommand{\Tpnpn}{\ensuremath{T_{\pi N,\pi N}}}
\newcommand{\Tpnppn}{\ensuremath{T_{\pi\pi N,\pi N}}}
\newcommand{\Tonpn}{\ensuremath{T_{\omega N,\pi N}}}
\newcommand{\tonpn}{\ensuremath{t_{\omega N,\pi N}}}
\newcommand{\tronpn}{\ensuremath{t^R_{\omega N,\pi N}}}
\newcommand{\Tongn}{\ensuremath{T_{\omega N,\gamma N}}}
\newcommand{\tongn}{\ensuremath{t_{\omega N,\gamma N}}}
\newcommand{\trongn}{\ensuremath{t^R_{\omega N,\gamma N}}}
\newcommand{\vmbgn}{\ensuremath{v_{MB,\gamma N}}}
\newcommand{\vpngn}{\ensuremath{v_{\pi N,\gamma N}}}
\newcommand{\vongn}{\ensuremath{v_{\omega N,\gamma N}}}
\newcommand{\vonpn}{\ensuremath{v_{\omega N,\pi N}}}
\newcommand{\vpnpn}{\ensuremath{v_{\pi N,\pi N}}}
\newcommand{\vonon}{\ensuremath{v_{\omega N,\omega N}}}
\newcommand{\vrngn}{\ensuremath{v_{\rho N,\gamma N}}}
\newcommand{\tjtmbmb}{\ensuremath{t^{JT}_{M'B',MB}}}
\newcommand{\tjlsmngn}{\ensuremath{t^{JT}_{L'S'M'N',\lag\lN T_{N,z}}}}
\newcommand{\tjlsmbgn}{\ensuremath{t^{JT}_{LSMB,\lag \lN T_{N,z}}}}
\newcommand{\vjlsmngn}{\ensuremath{v^{JT}_{L'S'M'N',\lag \lN T_{N,z}}}}
\newcommand{\vjlsmbgn}{\ensuremath{v^{JT}_{LSMB,\lag \lN T_{N,z}}}}
\newcommand{\tjlsmnmb}{\ensuremath{t^{JT}_{L'S'M'N',LSMB}}}
\newcommand{\Tjlsmbmb}{\ensuremath{T^{JT}_{LSMB,L'S'M'B'}}}
\newcommand{\tjlsmbmb}{\ensuremath{t^{JT}_{LSMB,L'S'M'B'}}}
\newcommand{\tjlsmnpn}{\ensuremath{t^{JT}_{L'S'M'N',\ell \pi N}}}
\newcommand{\tjlsmbpn}{\ensuremath{t^{JT}_{LSMB,\ell \pi N}}}
\newcommand{\vjlsmnpn}{\ensuremath{v^{JT}_{L'S'M'N',\ell \pi N}}}
\newcommand{\vjlsmnmb}{\ensuremath{v^{JT}_{L'S'M'N',LSMB}}}
\newcommand{\vjlsmbpn}{\ensuremath{v^{JT}_{LSMB,\ell \pi N}}}
\newcommand{\Tjlsmngn}{\ensuremath{t^{R,JT}_{L'S'M'N',\lag\lN T_{N,z}}}}
\newcommand{\Tjlsmbgn}{\ensuremath{t^{R,JT}_{LSMB,\lag \lN T_{N,z}}}}
\newcommand{\Tfjlsmbgn}{\ensuremath{T^{JT}_{LSMB,\lag \lN T_{N,z}}}}
\newcommand{\Tjlsmnmb}{\ensuremath{t^{R,JT}_{L'S'M'N',LSMB}}}
\newcommand{\Tjlsmnpn}{\ensuremath{t^{R,JT}_{L'S'M'N',\ell \pi N}}}
\newcommand{\Tjlsmbpn}{\ensuremath{t^{R,JT}_{LSMB,\ell \pi N}}}
\newcommand{\Gbjlsi}{\ensuremath{{\Gamma}^{JT}_{LSMB,N^*_i}}}
\newcommand{\Gbjlspi}{\ensuremath{{\Gamma}^{JT}_{L'S'M'B',N^*_i}}}
\newcommand{\Gjlsi}{\ensuremath{\overline{\Gamma}^{JT}_{LSMB,N^*_i}}}
\newcommand{\Gijls}{\ensuremath{\overline{\Gamma}^{JT}_{N^*_i,LSMB}}}
\newcommand{\Gbijls}{\ensuremath{{\Gamma}^{JT}_{N^*_i,LSMB}}}
\newcommand{\Gjpn}{\ensuremath{\overline{\Gamma}^{JT}_{N^*_j,\ell\pn}}}
\newcommand{\Gign}{\ensuremath{\overline{\Gamma}^{JT}_{N^*_i,\lag\lN T_{N,z}}}}
\newcommand{\Gbign}{\ensuremath{{\Gamma}^{JT}_{N^*_i,\lag\lN T_{N,z}}}}
\newcommand{\Gjlsj}{\ensuremath{\overline{\Gamma}^{JT}_{LSMB,N^*_j}}}
\newcommand{\Gjem}{\ensuremath{\overline{\Gamma}^{JT}_{N^*_j,\lag\lN T_{N,z}}}}
\newcommand{\Ljtlsmbn}{\ensuremath{\Lambda^{JT}_{N^*LSMB}}}
\newcommand{\Drij}{\ensuremath{\mathcal{D}^{-1}_{ij}}}
\newcommand{\Mbres}{\ensuremath{M^{(0)}_{N^*}}}
\newcommand{\Cjtnlsmb}{\ensuremath{C^{JT}_{N^*LSMB}}}
\newcommand{\Ljtnlsmb}{\ensuremath{\Lambda^{JT}_{N^*LSMB}}}
\newcommand{\knstar}{\ensuremath{k_{N^*}}}
\newcommand{\vonen}{\ensuremath{v_{\omega N,\eta N}}}
\newcommand{\vonpd}{\ensuremath{v_{\omega N,\pi\Delta}}}
\newcommand{\vonsn}{\ensuremath{v_{\omega N,\sigma N}}}
\newcommand{\vonrn}{\ensuremath{v_{\omega N,\rho N}}}
\newcommand{\gnon}{\ensuremath{\gamma N\to \omega N}}
\newcommand{\gnpn}{\ensuremath{\gamma N\to \pi N}}
\newcommand{\gnky}{\ensuremath{\gamma N\to KY}}
\newcommand{\enepn}{\ensuremath{e N\to e'\pi N}}
\newcommand{\gnen}{\ensuremath{\gamma N\to \eta N}}
\newcommand{\gpop}{\ensuremath{\gamma p\to \omega p}}
\newcommand{\gpep}{\ensuremath{\gamma p\to \eta p}}
\newcommand{\gpepp}{\ensuremath{\gamma p\to \eta' p}}
\newcommand{\gnten}{\ensuremath{\gamma n\to \eta n}}
\newcommand{\pzp}{\ensuremath{\pi^0 p}}
\newcommand{\ppln}{\ensuremath{\pi^+ n}}
\newcommand{\pmp}{\ensuremath{\pi^- p}}
\newcommand{\pzn}{\ensuremath{\pi^0 n}}
\newcommand{\gppzp}{\ensuremath{\gamma p\to \pi^0 p}}
\newcommand{\gpppn}{\ensuremath{\gamma p\to \pi^+ n}}
\newcommand{\gnpmp}{\ensuremath{\gamma n\to \pi^- p}}
\newcommand{\gnpzn}{\ensuremath{\gamma n\to \pi^0 n}}
\newcommand{\gppzep}{\ensuremath{\gamma p\to \pi^0 \eta p}}
\newcommand{\pnen}{\ensuremath{\pi N\to \eta N}}
\newcommand{\pnon}{\ensuremath{\pi N\to \omega N}}
\newcommand{\pnmb}{\ensuremath{\pi N\to MB}}
\newcommand{\gnmb}{\ensuremath{\gamma N\to M\!B}}
\newcommand{\onon}{\ensuremath{\omega N\to \omega N}}
\newcommand{\pmpon}{\ensuremath{\pi^- p\to \omega n}}
\newcommand{\pnpn}{\ensuremath{\pi N\to \pi N}}
\newcommand{\pnppn}{\ensuremath{\pi N\to\pi\pi N}}
\newcommand{\knkn}{\ensuremath{K N\to K N}}
\newcommand{\nnnn}{\ensuremath{N N\to N N}}
\newcommand{\pDpD}{\ensuremath{\pi D\to \pi D}}
\newcommand{\pDpp}{\ensuremath{\pi^+ D\to pp}}
\newcommand{\Gon}{\ensuremath{G_{0,\omega N}}}
\newcommand{\Gpn}{\ensuremath{G_{0,\pi N}}}
\newcommand{\rhomb}{\ensuremath{\rho_{MB}}}
\newcommand{\rhoon}{\ensuremath{\rho_{\omega N}}}
\newcommand{\rhopn}{\ensuremath{\rho_{\pi N}}}
\newcommand{\kon}{\ensuremath{k_{\omega N}}}
\newcommand{\kpn}{\ensuremath{k_{\pi N}}}
\newcommand{\Gmb}{\ensuremath{G_{0,MB}}}
\newcommand{\Tmbgn}{\ensuremath{T_{MB,\gamma N}}}
\newcommand{\vmbpgn}{\ensuremath{v_{M'B',\gamma N}}}
\newcommand{\pntpn}{\ensuremath{\pi N\!\to\!\pi N}}
\newcommand{\pnten}{\ensuremath{\pi N\!\to\!\eta N}}
\newcommand{\pnton}{\ensuremath{\pi N\!\to\!\omega N}}
\newcommand{\epos}{\ensuremath{\slashed{\epsilon}_{\lambda_\omega}}}
\newcommand{\epo}{\ensuremath{{\epsilon}_{\lambda_\omega}}}
\newcommand{\elevi}{\ensuremath{{\epsilon}_{\alpha\beta\gamma\delta}}}
\newcommand{\eps}{\ensuremath{\epsilon}}
\newcommand{\krho}{\ensuremath{\kappa_\rho}}
\newcommand{\komg}{\ensuremath{\kappa_\omega}}
\newcommand{\komga}{\ensuremath{\kappa^t_\omega}}
\newcommand{\doh}{\ensuremath{d^{(\half)}_{\lambda'\lambda}}}
\newcommand{\dohm}{\ensuremath{d^{(\half)}_{-\lambda,-\lambda'}}}
\newcommand{\dohmo}{\ensuremath{d^{(\half)}_{\lambda',-\half}}}
\newcommand{\dohpo}{\ensuremath{d^{(\half)}_{\lambda',+\half}}}
\newcommand{\Lor}[2]{\ensuremath{\Lambda^{#1}_{\hphantom{#1}{#2}}}}
\newcommand{\ILor}[2]{\ensuremath{\Lambda_{#1}^{\hphantom{#1}{#2}}}}
\newcommand{\LorT}[2]{\ensuremath{[\Lambda^T]^{#1}_{\hphantom{#1}{#2}}}}
\newcommand{\dsdo}{\ensuremath{{\frac{d\sigma}{d\Omega}}}}
\newcommand{\dsdol}{\ensuremath{{\frac{d\sigma}{d\Omega_l}}}}
\newcommand{\dsdolo}{\ensuremath{{\frac{d\sigma}{d\Omega_{l,1}}}}}
\newcommand{\dsdoc}{\ensuremath{{\frac{d\sigma}{d\Omega_c}}}}
\newcommand{\dsdon}{\ensuremath{{{d\sigma}/{d\Omega}}}}
\newcommand{\dspdo}{\ensuremath{{\frac{d\sigma_\pi}{d\Omega}}}}
\newcommand{\dsgdo}{\ensuremath{{\frac{d\sigma_\gamma}{d\Omega}}}}
\newcommand{\chipd}{\ensuremath{\chi^2/N_d}}
\newcommand{\chipda}{\ensuremath{\chi^2(\alpha)/N_d}}
\newcommand{\bpop}{\ensuremath{\bvec{p}'_1}}
\newcommand{\bptp}{\ensuremath{\bvec{p}'_2}}
\newcommand{\bpip}{\ensuremath{\bvec{p}'_i}}
\newcommand{\bpo}{\ensuremath{\bvec{p}_1}}
\newcommand{\bpt}{\ensuremath{\bvec{p}_2}}
\newcommand{\bpi}{\ensuremath{\bvec{p}_i}}
\newcommand{\bqo}{\ensuremath{\bvec{q}_1}}
\newcommand{\bqt}{\ensuremath{\bvec{q}_2}}
\newcommand{\bqi}{\ensuremath{\bvec{q}_i}}
\newcommand{\bQ}{\ensuremath{\bvec{Q}}}
\newcommand{\bq}{\ensuremath{\bvec{q}}}
\newcommand{\ketq}{\ensuremath{\ket{\bqo,\bqt}}}
\newcommand{\ketqc}{\ensuremath{\ket{\bQ,\bq}}}
\newcommand{\bP}{\ensuremath{\bvec{P}}}
\newcommand{\bPp}{\ensuremath{\bvec{P}'}}
\newcommand{\bpr}{\ensuremath{\bvec{p}}}
\newcommand{\bprp}{\ensuremath{\bvec{p}'}}
\newcommand{\ketPsiq}{\ensuremath{\ket{\Psi_{\bq}^{(\pm)}}}}
\newcommand{\ketPsiqQ}{\ensuremath{\ket{\Psi_{\bQ,\bq}^{(\pm)}}}}
\newcommand{\Ld}{\ensuremath{\mathcal{L}}}
\newcommand{\ps}{\mbox{ps}}
\newcommand{\fndp}{f_{N\Delta\pi}}
\newcommand{\fndr}{f_{N\Delta\rho}}
\newcommand{\said}{{\sc said}}
\newcommand{\ret}{\ensuremath{\langle{\tt ret}\rangle}}
\newcommand{\ddf}[1]{\ensuremath{\delta^{(#1)}}}
\newcommand{\Tpp}{\ensuremath{T_{\pi\pi}}}
\newcommand{\Kpp}{\ensuremath{K_{\pi\pi}}}
\newcommand{\Tpe}{\ensuremath{T_{\pi\eta}}}
\newcommand{\Kpe}{\ensuremath{K_{\pi\eta}}}
\newcommand{\Tep}{\ensuremath{T_{\eta\pi}}}
\newcommand{\Kep}{\ensuremath{K_{\eta\pi}}}
\newcommand{\Tee}{\ensuremath{T_{\eta\eta}}}
\newcommand{\Kee}{\ensuremath{K_{\eta\eta}}}
\newcommand{\Tpig}{\ensuremath{T_{\pi\gamma}}}
\newcommand{\Kpig}{\ensuremath{K_{\pi\gamma}}}
\newcommand{\oKpig}{\ensuremath{\overline{K}_{\pi\gamma}}}
\newcommand{\tKpig}{\ensuremath{\tilde{K}_{\pi\gamma}}}
\newcommand{\Teg}{\ensuremath{T_{\eta\gamma}}}
\newcommand{\Keg}{\ensuremath{K_{\eta\gamma}}}
\newcommand{\Kab}{\ensuremath{K_{\alpha\beta}}}
\newcommand{\R}{\ensuremath{\mathbb{R}}}
\newcommand{\C}{\ensuremath{\mathbb{C}}}
\newcommand{\Ezp}{\ensuremath{E^{\pi}_{0+}}}
\newcommand{\Eze}{\ensuremath{E^{\eta}_{0+}}}
\newcommand{\Ga}{\ensuremath{\Gamma_\alpha}}
\newcommand{\Gb}{\ensuremath{\Gamma_\beta}}
\newcommand{\RH}{\ensuremath{\mathcal{R}\!\!-\!\!\mathcal{H}}}
\newcommand{\calT}{\mathcal{T}}
\newcommand{\maid}{{\sc maid}}
\newcommand{\Kbar}{\ensuremath{\overline{K}}}
\newcommand{\zbar}{\ensuremath{\overline{z}}}
\newcommand{\kbar}{\ensuremath{\overline{k}}}
\newcommand{\dom}{\ensuremath{\mathcal{D}}}
\newcommand{\domi}[1]{\ensuremath{\mathcal{D}_{#1}}}
\newcommand{\nubar}{\ensuremath{\overline{\nu}}}
\newcommand{\pbar}{\ensuremath{\overline{p}}}
\newcommand{\Nab}{\ensuremath{N_{\alpha\beta}}}
\newcommand{\Nee}{\ensuremath{N_{\eta\eta}}}
\newcommand{\dth}[1]{\delta^{(3)}(#1)}
\newcommand{\dfo}[1]{\delta^{(4)}(#1)}
\newcommand{\intd}[1]{\int\!d{#1}\,}
\newcommand{\intdl}[3]{\int_{#2}^{#3}\!d{#1}\,}
\newcommand{\intD}[1]{\int\!\mathscr{D}{#1}\,}
\newcommand{\intk}{\int\!\!\frac{d^3\! k}{(2\pi)^3}}
\newcommand{\intkg}{\int\!\!{d^3\! k_\gamma}}
\newcommand{\intks}{\int\!\!{d^3\! k_\sigma}}
\newcommand{\nch}{\ensuremath{N_{\mbox{ch}}}}
\newcommand{\nc}{\ensuremath{N_{ch}}}
\newcommand{\re}{\ensuremath{\mbox{Re }\!}}
\newcommand{\im}{\ensuremath{\mbox{Im }\!}}
\newcommand{\EetaS}{\ensuremath{E^\eta_{0+}}}
\newcommand{\EpiS}{\ensuremath{E^\pi_{0+}}}
\newcommand{\tobull}{\ensuremath{\to}}
\newcommand{\Kcm}{\ensuremath{K_{CM}}}
\newcommand{\lra}{\ensuremath{\leftrightarrow}}
\newcommand{\avg}[1]{\ensuremath{\langle{#1}\rangle}}
\newcommand{\Chi}{\ensuremath{\text{Chi}}}
\newcommand{\Shi}{\ensuremath{\text{Shi}}}

\newcommand{\gn}{\ensuremath{\gamma N}}
\newcommand{\gp}{\ensuremath{\gamma p}}
\newcommand{\geta}{\ensuremath{\gamma \eta}}
\newcommand{\pp}{\ensuremath{pp}}
\newcommand{\pn}{\ensuremath{\pi N}}
\newcommand{\pide}{\ensuremath{\pi d}}
\newcommand{\en}{\ensuremath{\eta N}}
\newcommand{\epn}{\ensuremath{\eta' N}}
\newcommand{\pD}{\ensuremath{\pi \Delta}}
\newcommand{\sn}{\ensuremath{\sigma N}}
\newcommand{\rn}{\ensuremath{\rho N}}
\newcommand{\on}{\ensuremath{\omega N}}
\newcommand{\ppn}{\ensuremath{\pi\pi N}}
\newcommand{\pipi}{\ensuremath{\pi\pi}}
\newcommand{\kn}{\ensuremath{KN}}
\newcommand{\ky}{\ensuremath{KY}}
\newcommand{\kl}{\ensuremath{K\Lambda}}
\newcommand{\ks}{\ensuremath{K\Sigma}}
\newcommand{\bn}{\ensuremath{eN}}
\newcommand{\pR}{\ensuremath{\pi N^*}}
\newcommand{\bpn}{\ensuremath{e\pi N}}
\newcommand{\fpo}{\ensuremath{5\oplus 1}}
\newcommand{\faoe}{{\sc FA08}}
\newcommand{\fpoe}{{\sc FP08}}
\newcommand{\fsoe}{{\sc FS08}}
\newcommand{\psic}{\ensuremath{\psi_{n\kappa jm}}}
\newcommand{\hi}[1]{\ensuremath{H_I(t_{#1})}}
\newcommand{\dcp}{\ensuremath{\mathcal{D}}}
\newcommand{\pptopp}{\ensuremath{\pipi\to\pipi}}
\newcommand{\pntoppn}{\ensuremath{\pn\to\ppn}}
\newcommand{\ovlt}{\ensuremath{\overline{t}}}
\newcommand{\smat}{\ensuremath{e^{-i\int_{-\infty}^\infty\!dt\,H_I(t)}}}
\newcommand{\pder}[2][]{\frac{\partial#1}{\partial#2}}
\newcommand{\der}[2][]{\frac{d#1}{d#2}}

\newcommand{\Rhmat}{$R$-matrix}
\newcommand{\Rmat}{$R$ matrix}
\newcommand{\svb}{\ensuremath{\langle\sigma v\rangle}}
\newcommand{\redG}{\ensuremath{\tilde{G}}}

\newcommand{\mHn}{\ensuremath{^1\mbox{H}}}
\newcommand{\Hii}{\ensuremath{^2\mbox{H}}\xspace}
\newcommand{\Hn}{H}
\newcommand{\mde}{\ensuremath{d}}
\newcommand{\de}{d}
\newcommand{\mtriton}{\ensuremath{t}}
\newcommand{\triton}{t}
\newcommand{\mHiii}{\ensuremath{^3\!\text{H}}}
\newcommand{\Hiii}{$^3$H\xspace}
\newcommand{\mHeiii}{\ensuremath{^3\!\text{He}}}
\newcommand{\Heiii}{$^{3}$He\xspace}
\newcommand{\mHeiv}{\ensuremath{^4\!\text{He}}}
\newcommand{\Heiv}{$^4$He\xspace}
\newcommand{\jHeiv}{$^4$He}
\newcommand{\mHev}{\ensuremath{^5\!H\!e}}
\newcommand{\Hev}{\ensuremath{^5}He}
\newcommand{\mHevi}{\ensuremath{^6\!He}}
\newcommand{\Hevi}{\ensuremath{^6}He}
\newcommand{\mLivi}{\ensuremath{^6\!Li}}
\newcommand{\Livi}{\ensuremath{^6}Li}
\newcommand{\mLivii}{\ensuremath{^7\!Li}\xspace}
\newcommand{\Livii}{$^7$Li\xspace}
\newcommand{\jLivii}{$^7$Li}
\newcommand{\mBevii}{\ensuremath{^7\!Be}}
\newcommand{\Bevii}{${^7}$Be\xspace}
\newcommand{\mBeviii}{\ensuremath{^8\!Be}}
\newcommand{\Beviii}{\ensuremath{^8}Be}
\newcommand{\mBeix}{\ensuremath{^9\!Be}}
\newcommand{\Beix}{\ensuremath{^9}Be}
\newcommand{\mBix}{\ensuremath{^9\!B}}
\newcommand{\Bix}{\ensuremath{^9}B}
\newcommand{\mBx}{\ensuremath{^{10}\!B}}
\newcommand{\Bx}{\ensuremath{^{10}}B}
\newcommand{\mCxiii}{\ensuremath{^{13}\!C}}
\newcommand{\Cxiii}{\ensuremath{^{13}}C}
\newcommand{\mCxiv}{\ensuremath{^{14}\!C}}
\newcommand{\Cxiv}{\ensuremath{^{14}}C}
\newcommand{\Oxvii}{\ensuremath{^{17}}O}

\newcommand{\EDA}{{\tt EDA}}
\newcommand{\ttinp}{{\tt input}}
\newcommand{\ttout}{{\tt output}}
\newcommand{\ttdat}{{\tt data}}
\newcommand{\ttpar}{{\tt par}}
\newcommand{\ttnam}{{\tt namelst}}
\newcommand{\tthol}{{\tt hold}}
\newcommand{\tthne}{{\tt hnew}}
\newcommand{\tteda}{{\tt eda}}
\newcommand{\ttsou}{{\tt sout}}
\newcommand{\ttpan}{{\tt parnew}}
\newcommand{\tturd}{{\tt urd}}
\newcommand{\ttfen}{{\tt fengy}}
\newcommand{\ttfet}{{\tt fengy2}}
\newcommand{\ttplo}{{\tt plotf}}
\newcommand{\ttpao}{{\tt par1r}}
\newcommand{\ttfdb}{{\tt fdbdu}}
\newcommand{\unk}{{\bf unk}}
\newcommand{\ttchr}{{\tt changri}}

\newcommand{\itPFP}{\textit{Physics for Future Presidents}}
\newcommand{\itaPFP}{\textit{PFP}}
\newcommand{\prle}{\textit{PR}\textbf{97}}
\newcommand{\lcd}{---\;}

\newcommand{\cnb}{C$\nu$B}
\newcommand{\nue}{\ensuremath{\nu_e}\xspace}
\newcommand{\bnue}{\ensuremath{\overline\nu_e}\xspace}
\newcommand{\numu}{\ensuremath{\nu_\mu}\xspace}
\newcommand{\bnumu}{\ensuremath{\overline\nu_\mu}\xspace}
\newcommand{\nut}{\ensuremath{\nu_\tau}\xspace}
\newcommand{\bnut}{\ensuremath{\overline\nu_\tau}\xspace}
\newcommand{\Gcs}[2]{\ensuremath{\Gamma\indices{^{#1}_{#2}}}}
\newcommand{\Gcst}[2]{\ensuremath{{\tilde\Gamma}\indices{^{#1}_{#2}}}}
\newcommand{\flrw}{Friedmann-Lema\^{i}tre-Robertson-Walker}
\newcommand{\mpl}{\ensuremath{m_{pl}}}
\newcommand{\gnuen}{\ensuremath{\gamma\nu eN}}
\newcommand{\cdrv}{\ensuremath{\frac{d\hphantom{t}}{dt}}}
\newcommand{\neff}{\ensuremath{N_{\mbox{\scriptsize eff}}}}
\newcommand{\lcdm}{\ensuremath{\Lambda}CDM}
\newcommand{\heii}{\mbox{\tiny He {\sc ii}}}

\newcommand{\km}{\mbox{km}}
\newcommand{\me}{\mbox{m}}
\newcommand{\cm}{\mbox{cm}}
\newcommand{\mm}{\mbox{mm}}
\newcommand{\s}{\mbox{s}}
\newcommand{\kg}{\mbox{kg}}
\newcommand{\g}{\mbox{g}}
\newcommand{\Mpc}{\mbox{Mpc}}
\newcommand{\GeV}{\mbox{GeV}}

\newcommand{\neffz}{\ensuremath{n^{(0)}_\text{eff}}}
\newcommand{\zpf}{\ensuremath{\mathcal{Z}}}

\title{Effect of collisions on neutrino flavor inhomogeneity
in a dense neutrino gas} 

\author{Vincenzo Cirigliano}
\ead{cirigliano@lanl.gov}
\author{Mark W.\ Paris}
\ead{mparis@lanl.gov}
\author{Shashank Shalgar}
\ead{shashank@lanl.gov}
\address{Theoretical Division, Los Alamos National Laboratory, 
Los Alamos, NM 87545, USA}

\begin{abstract}

We investigate the stability, with respect to spatial inhomogeneity,
of a two-dimensional dense neutrino gas. The system exhibits growth of
seed inhomogeneity due to nonlinear coherent neutrino
self-interactions. In the absence of incoherent collisional effects,
we observe a dependence of this instability growth rate on the
neutrino mass spectrum:  the normal neutrino mass hierarchy exhibits
spatial instability over a larger range of neutrino number density
compared to that of the inverted case. We further consider the effect
of elastic incoherent collisions of the neutrinos with a static
background of heavy, nucleon-like scatterers. At small scales, the
growth of flavor instability can be suppressed by collisions.  At
large length scales we find, perhaps surprisingly, that for inverted
neutrino mass hierarchy incoherent collisions fail to suppress flavor
instabilities, independent of the coupling strength.

\end{abstract}

\begin{keyword}
neutrino oscillations \sep dense neutrino medium
\end{keyword}

\maketitle

\section{Introduction} 
\label{sec:intro}

Several recent studies, concerned primarily  with  anisotropic
astrophysical environments,  have demonstrated that the nonlinear
neutrino self-coupling results in spatial
instabilities~\cite{Duan:2014gfa,Abbar:2015mca,Mirizzi:2015fva,
Chakraborty:2015tfa}.  These works indicate the importance of taking
into account deviations from spherical symmetry in describing the
evolution of, for example, the supernova neutrino flavor field.
Previous studies of the neutrino
evolution~\cite{Dolgov:1997mb,Dolgov:2002ab,Mangano:2005cc,
Grohs:2015eua,Grohs:2015tfy} in the early universe generally assume
homogeneity and isotropy obtains at all times from the epochs of weak
equilibrium through neutrino decoupling and Big Bang nucleosynthesis
(BBN). The objective of the present study is to address the validity
of this assumption in an exploratory calculation.  To this end,  in
this Letter we study the stability of a two-dimensional, two-flavor
dense neutrino gas with respect to the growth of seed spatial
inhomogeneity driven by neutrino self-coupling. In this model, we
also investigate the impact of incoherent elastic collisions on the
stability properties.  We find, perhaps surprisingly, that elastic
scattering-angle dependent collisions are not always effective at
damping flavor instabilities.

In extreme environments with large neutrino number densities, such as
the interior of a core-collapse supernovae (CCSN) and in the early
universe,  the nonlinearity associated with neutrino self-interactions
lead to an array of interesting effects beyond the ``standard''
MSW~\cite{Wolfenstein:1977ue,Mikheev:1986gs}.  These effects, which
have their origin in the non-linear character of the neutrino
evolution equations, include collective oscillations in an idealized
model of supernova~\cite{Pantaleone:1994ns,
Duan:2005cp,Duan:2006an,Duan:2006jv,Duan:2007mv,Duan:2008za,
Raffelt:2007xt,EstebanPretel:2007bz,EstebanPretel:2008ni,
Raffelt:2008hr, Dasgupta:2010ae}, matter-neutrino resonant
behavior~\cite{Malkus:2014iqa,Zhu:2016mwa,Wu:2015fga}, spontaneous
breaking of the axial symmetry~\cite{Raffelt:2013rqa,Mirizzi:2013wda},
and emergence of spatial
inhomogeneity~\cite{Duan:2014gfa,Abbar:2015mca}, to cite a few
examples.

In this context, the early universe is particularly interesting to study since    
non-equilibrium and neutrino flavor oscillation effects may be important for 
precision cosmological probes,  such as BBN.   
In the early universe neutrino oscillations can be  induced by 
flavor and/or  lepton asymmetries.

Studies of neutrino weak decoupling in a Boltzmann equation
approach~\cite{Dolgov:1997mb,Mangano:2001iu,Grohs:2015tfy}  suggest
that neutrino spectra  deviate from equilibrium Fermi-Dirac
distributions at the percent level and display flavor dependence (\nue
versus  \numu/\nut) at the  level of a few
percent~\cite{Dolgov:1997mb,Grohs:2015tfy}.  The flavor asymmetry
triggers flavor oscillations and a full analysis of this problem is
still lacking. Significant progress, however, has been made in
Refs.~\cite{Dolgov:2002ab,Mangano:2005cc,deSalas:2016ztq}.

Additionally, and complementary to the question of neutrino flavor
asymmetry,   neutrino oscillations play a significant role in the
presence of a lepton number asymmetry, namely a difference in  the
spectra for neutrinos and antineutrinos  (e.g. $\nu_{e}$ and
$\bar{\nu}_{e}$).
An electron lepton asymmetry large enough for collective neutrino
oscillations to be important in the  early universe can be generated
in  several well  motivated
models~\cite{Harvey:1981cu,Foot:1995qk,Shi:1996ic,Casas:1997gx,
MarchRussell:1999ig,Kawasaki:2002hq,Yamaguchi:2002vw,Shaposhnikov:2008pf,
Gu:2010dg}.    Moreover, a recent study~\cite{Johns:2016enc},
examining a range of lepton number flavor asymmetries over six orders
of magnitude, identified a variety of qualitatively distinct regimes
and  non-trivial behavior at very small lepton asymmetry.

The departure point for this study, in light of  the complex and
sundry array of behavior driven by the nonlinear neutrino
self-interaction, is the question of spatial inhomogeneity. We ask the
question of whether the assumption of spatial homogeneity and
isotropy, unexamined in previous works on the flavor field evolution
in the early universe~\cite{Dolgov:2002ab,Mangano:2005cc}, is
sustained in the presence of the nonlinear neutrino self-interaction.
Our finding is that, for a two-dimensional, two-flavor
exploratory model of a neutrino gas, the system is indeed unstable
with respect to spatial inhomogeneity. Further, we ask the question of
whether incoherent collisions, with a static background field of
elastic scatterers, can serve to drive the neutrino field back to
homogeneity. We investigate inhomogeneity for both normal and
inverted neutrino mass hierarchies.  We find that for the inverted
neutrino spectrum such collisions  (irrespective of the coupling
strength) are incapable of stabilizing the system for a significant
range of length scales.

Instability does not imply, of course, the persistence of
inhomogeneity for all times. Eventually, incoherent collisions may in
fact drive the system toward homogeneity. In this case, the associated
transient inhomogeneity is associated with entropy generation. And, if
sufficiently large, this added entropy may effect the evolution of the
light element nucleosynthesis in early universe. This is a question
that can only be addressed with detailed numerical simulations; such a
code is currently under
development~\cite{Grohs:2015tfy,Grohs:2016vef,Grohs:2016cuu}.

Our consideration of a simplified model of the early universe is
driven mainly by pragmatic concerns.  The rich treasure of effects
that are closely related to nonlinearity in the collective neutrino
oscillations makes prediction difficult for realistic physical
systems. Any such prediction involving collective neutrino
oscillations in, for example, the interior of the proto-neutron star
formed in CCSN requires the solution of the neutrino-flavor density
matrix in a large dimensional space,\footnote{The general case of
inhomogeneous and anisotropic environments, relevant for supernovae,
corresponds to an equation of motion for the neutrino density matrix
in a space of seven dimensions corresponding to three spatial, three
momentum coordinates and time.} which is beyond the capability of even
modern large-scale, multiprocessor computational platforms. However,
it is possible to solve these equations for simplified models that
assume deviations from highly symmetrical spatial geometries may be 
neglected thereby reducing the dimensionality of the
solution space. It is hoped that, with this approach, insight may be
gained into the development of methods that will allow the solution of
the full problem; or, at least, to obtain results of phenomenological
relevance. 

This Letter is organized as follows. 
In Sec.~\ref{formalism}, we detail the model of neutrino oscillations
employed in this exploratory study of neutrino flavor stability in the
early universe. Section \ref{collision} discusses the neutrino-nucleus
model of the elastic, angle-dependent collision term adopted for the
model. In Sec.~\ref{linearstability}, we perform linear stability
analysis of the model and present results in Sec.~\ref{sec:results}.
Finally, in Sec.~\ref{sec:concl}, we explore the implications and
limitations of the present study in the context of more realistic
models. In particular, we comment on effects that inelastic
contributions to the collisions may have on collective neutrino
oscillations. 

\section{Model setup} 
\label{formalism}

The simplification of the spacetime and four-momentum dependence of
the equations of motion of a dense neutrino gas in conditions similar
to that of the early universe is effected in three stages:
simplification of the geometry; restriction to two neutrino flavors;
and approximation of the collision terms. The equations of motion that
govern the evolution of the neutrino density matrix in the early
universe depend on spacetime position and
the neutrino three-momentum and energy. The equations are simplified
by reducing the  spatial dimensions of the universe from three to two.   
Since we are performing the linearized
stability analysis at a given instant in the evolution of the system,
we may further neglect the spacetime curvature, without loss of
generality.   Our assumption of elastic collisions with a background static array of
nucleon-like scatterers allows the further assumption that neutrino
energies are decoupled; energy transport is neglected. Thus we
consider the neutrinos to be mono-energetic. This reduces the momentum
dependence to that of a single, angular coordinate $\theta$. The
resulting equations of motion are  four-dimensional -- two spatial
dimensions, one momentum direction and time.
We work with two active neutrino flavors, instead of three. 

The purpose of restricting the dimensionality of spacetime to a planar
spatial surface is to limit the size of the space of Fourier modes
(see Eq.~\eqref{mom:exp} below) that must be considered in the
stability analysis. In fact, we expect that this approximation is not
too severe in the sense that, all other approximations taken equally,
a treatment considering the universe as three spatial dimensions would
simply result in a larger number of Fourier modes. The stability
behaviors  of at least some of the modes in this case would be
similar to those of the two-dimensional case since the spatial mode
wave number couples explicitly only to the local velocity of the
neutrino field.  The two-flavor approximation also reduces the size of
the space in which the linearized stability analysis is performed.
The purpose of restricting the collision term to have only angular
dependence in elastic scatterings of the type $\nu_\alpha N \to
\nu_\alpha N$, where the field $N$ is represented by an infinitely
massive, immobile object, is to allow consideration of just a single
neutrino energy, which is conserved by this process.  With these
simplifications the universe can be viewed as a square of length $L$
on each side; we may take periodic boundary conditions without loss of
generality.

The dense neutrino gas is described in terms of the density matrix in
flavor space with elements $f_{\alpha \beta}(x,p)$, where $x$ and $p$
are position and momentum four-vectors, formally related to the Wigner
transform of the neutrino correlation function  $\langle \nu_\alpha
(x) \bar{\nu}_\beta (y)\rangle$ in
medium~\cite{Cirigliano:2014aoa,Vlasenko:2013fja,Blaschke:2016xxt}.  Factorization of the total
population (flux) $n(\vec{p})$ in momentum bin $\vec{p}$, a scalar
quantity, from the correlation function yields the $2\times 2$ density
matrices $\rho^{\theta}(\vec{x},t)$ and
$\bar{\rho}^{\theta}(\vec{x},t)$ for neutrinos and anti-neutrinos
respectively, which describe the flavor content of the system; they
satisfy ${\rm Tr } \rho^\theta  = 1 = {\rm Tr } \bar \rho^\theta$. Here
$\vec{x}$ is a two-dimensional vector specifying the position,
$\theta$ is the polar angle characterizing the direction of the
neutrino momentum $\vec{p}$, with respect to a given, arbitrary
direction (taken as the $y$-axis);  $t$ is the time. The diagonal
components of the density matrix represent the relative populations
for the two neutrino flavors, denoted \nue and \numu. The off-diagonal
terms express correlations between the two flavors.
The spacetime evolution equations of the density matrices are
the quantum kinetic equations (QKEs)~\cite{Sigl:1992fn,Raffelt:1992uj,McKellar:1992ja,
Enqvist:1990ad,Strack:2005ux,Volpe:2013jgr,Volpe:2015rla,Vlasenko:2013fja,
Zhang:2013lka,Cirigliano:2014aoa,Serreau:2014cfa,Blaschke:2016xxt}, which are given, in the limit that we neglect incoherent
scattering, by:
\begin{align}
\label{eqn:dm_eom}
i\left(\frac{\partial}{\partial t}
+\vec{v}\cdot\vec{\nabla}
+\vec{F}\cdot\nabla_{p}\right)\rho^{\theta}(\vec{x},t) 
&= [\mathsf{H},\rho^{\theta}(\vec{x},t)], \cr
i\left(\frac{\partial}{\partial t}
+\vec{v}\cdot\vec{\nabla}
+\vec{F}\cdot\nabla_{p}\right)\bar{\rho}^{\theta}(\vec{x},t) 
&= [\bar{\mathsf{H}},\bar{\rho}^{\theta}(\vec{x},t)],
\end{align}
where $\vec{v} = \dot{\vec{x}}$, $\vec{F} = \dot{\vec{p}}$.
Here the Hamiltonian $\mathsf{H}$ for neutrino evolution and that for
antineutrino evolution $\bar{\mathsf{H}}$ is written as a sum of terms
\begin{align}
\mathsf{H} &=\mathsf{H}_{0} + \mathsf{H}_{\nu\nu}, \nonumber \\
\bar{\mathsf{H}} &=-\mathsf{H}_{0} + \mathsf{H}_{\nu\nu},
\label{eom}
\end{align}
where $\mathsf{H}_{0}$ is that part of the time-evolution operator
independent of the density matrix, which is comprised of the vacuum and
matter parts of the Hamiltonian. We ignore the term
$\vec{F}\cdot\nabla_{p}$, which contributes when external forces, such
as those induced by the Hubble expansion, are present.  It does not
contribute significantly in the early universe to the linearized
stability analysis.  The neutrino-neutrino evolution part of the
Hamiltonian, which depends linearly on the density matrix, is denoted by
$\mathsf{H}_{\nu\nu}$. These components are given explicitly as:
\begin{align}
\mathsf{H}_{0} &= \frac{1}{2}
\begin{pmatrix} -\omega & 0 \cr 0 & \omega \end{pmatrix} \\
\label{eqn:Hnn}
\mathsf{H}_{\nu\nu} &= \mu 
\int_{0}^{2\pi} d\theta^{\prime} 
\left( \rho^{\theta^{\prime}}(t,\vec{x})
      -\alpha\bar{\rho}^{\theta^{\prime}}(t,\vec{x})\right)
      (1-\cos(\theta-\theta^{\prime})).
\end{align}
Here  $\mu  \propto  G_F n_{\nu} $  denotes the effective
self-coupling strength, proportional to the neutrino number density, 
and $\alpha$  denotes the ratio $n_{\bar{\nu}}/n_{\nu}$.  The
linear, vacuum/matter term $\mathsf{H}_0$ is written for zero vacuum
mixing angle  ($\theta_V=0$) since $\theta_V$  does not alter the linearized stability analysis
we perform in Sec.~\ref{linearstability}.\footnote{Including  $\theta_V \neq 0$  would  have two effects:
(i) it would rescale the vacuum/matter frequency $\omega \to  \omega \times \cos 2 \theta_V$; (ii) it would 
introduce a source term $\sim  \omega \times \sin 2 \theta_V$ in the evolution of the homogeneous perturbations, 
which would lead to a linear growth with time, thus not impacting the stability analysis concerned with exponential growth.
Physically, the effect (ii)   corresponds to the onset of vacuum oscillations.} 
We have carried out the integration over energy assuming a $\delta$-function distribution for energy. 
Unlike the case of supernova there is no matter suppression of flavor
oscillations in a homogeneous and isotropic gas. 
We note, for later
discussion, that the normal hierarchy case corresponds to $\omega>0$
and $\omega<0$ is the inverted case for this equivalent two-flavor
system; we may refer to $\omega = \Delta m^2 / (2 E)$ as the `scaled' vacuum frequency.

We are now in a position to investigate the role of collective neutrino
oscillations in the early universe and whether the equations of motion
of the neutrino density matrix are stable with respect to seed
inhomogeneous perturbations. The analysis is carried out through the
decomposition of the equations of motion of the density matrices
[Eq.~\eqref{eqn:dm_eom}] in terms of their Fourier modes,
\begin{align}
\rho^{\theta}(t,\vec{x}) 
&= \sum_{\vec{k}}\rho^{\theta}_{\vec{k}}(t) \exp(i \vec{k}\cdot \vec{x}).
\label{mom:exp}
\end{align}
Here the sum is over $k_i = 2\pi n_i / L$ for $n_i \in \mathbb{Z}$ and
$i=x,y$. We note that previous studies have exclusively considered
only the homogeneous mode $\vec{k}=(0,0)$ (in two spatial
dimensions); all other modes $\vec{k}\ne\vec{0}$ are inhomogeneous. In
the remainder of the present work we write the equations of motion
[Eq.~\eqref{eqn:dm_eom}] in terms of the Fourier modes
$\rho^{\theta}_{\vec{k}}$. The objective of the linearized stability
analysis is to identify those modes that grow exponentially with time
and are therefore unstable to seed perturbations. Substituting 
Eq.~\eqref{mom:exp} in Eq.~\eqref{eom} and taking the projection to
the mode $\vec{k}$ gives
\begin{align}
i\frac{\partial}{\partial t} \rho^{\theta}_{\vec{k}}(t)
&= \vec{v}\cdot \vec{k} \rho^{\theta}_{\vec{k}}(t) 
+ \left[\mathsf{H}_{0},\rho^{\theta}_{\vec{k}}(t)\right] \nonumber \\
&+ \mu \sum_{k^{\prime}}\int_{0}^{2\pi} d\theta^{\prime}
  \left[(\rho^{\theta^{\prime}}_{\vec{k}-\vec{k^{\prime}}}(t) 
  -\alpha\bar{\rho}^{\theta^{\prime}}_{\vec{k}-\vec{k^{\prime}}}(t)),  
  \rho^{\theta}_{\vec{k^{\prime}}}(t) \right] \nonumber \\
  &\times (1-\cos(\theta-\theta^{\prime})).
\label{eom:mom2}
\end{align}
Pairs of modes that appear in the last term of Eq.~\eqref{eom:mom2}
sum to $\vec{k}$. 
Before considering the stability properties of the above collisionless
equations in Sec.~\ref{linearstability} we first discuss, in the 
following section, the general form of the
collision term and the elastic, angular dependent approximation we
make for the purposes of the present study.

\section{The collision term}
\label{collision}

The  QKEs presented  in the previous section assumed that the
neutrinos experience only the effects of forward scattering from other
neutrinos. We must, however, take into account the effect of
incoherent, direction changing
scattering off other neutrinos, electrons and nucleons present in the  
early
universe~\cite{Sigl:1992fn,Dolgov:1997mb,Grohs:2015tfy,Grohs:2016vef,Grohs:2016cuu}.
In this section, we
briefly review the incorporation of momentum changing scattering to
the neutrino flavor evolution. We continue to   assume the
reduced two-dimensional configuration space with  monoenergetic
neutrinos,   which greatly simplifies the scattering matrices. 
The stipulation that neutrinos remain monoenergetic
throughout the scattering requires that the scattering  
kernel only changes the direction of  the momentum and not magnitude.
We note that energy transport among neutrinos in  scattering events 
is expected to contribute to  additional suppression of the collective effects (see discussion in Sec.~\ref{sec:concl}).

Previous studies have included the effect of incoherent scattering 
on neutrino oscillations in an approximate   manner. Some references have 
included an off-diagonal damping term with Boltzmann collision integrals 
in the diagonal terms~\cite{Mangano:2011ip,Castorina:2012md,Pastor:2008ti,
Mangano:2010ei}. (A recent study~\cite{deSalas:2016ztq} has considered
collision terms off-diagonal in flavor but neglected terms where
neutrinos incoherently scatter from neutrinos and antineutrinos.) These 
treatments, which assume homogeneity and isotropy, have not considered 
the dependence of their results on the neutrino mass hierarchy. The
mass hierarchy, however, plays a role when inhomogeneity and anisotropy 
are present.

In order to calculate the effect of collisions on collective neutrino
oscillations we use the formalism developed
in~\cite{Blaschke:2016xxt}, albeit with a modified notation. The
general form for the rate of change of the density matrix due to
incoherent collisions is written as:
\begin{align}
\label{coll:nuc}
\frac{d}{dt}\rho^{\theta}(\vec{x},t) 
\Big\vert_{\rm coll} 
= &-\frac{1}{2}\left\{\Pi^{\textrm{loss}},\rho^{\theta}(\vec{x},t)\right\}
\nonumber \\
& +\frac{1}{2}\left\{\Pi^{\textrm{gain}},\left(1-\rho^{\theta}(\vec{x},t)\right)\right\}
.
\end{align}
The first term on the right-hand side encodes the flux loss for
neutrinos traveling in the $\theta$ direction and the second term
encodes the flux gain for a neutrino being scattered into the
direction $\theta$. We note that both $\rho$ and $\Pi^{\rm gain/loss} [\rho]$ are $2\times 2$
matrices and gain and loss terms are expressed in term of
anti-commutators denoted by $\left\{\ ,\ \right\}$.  The factor
$\left(1-\rho^{\theta}\right)$ in the second term is due to
Pauli-blocking, a consequence of the fermionic nature of the
neutrinos. Qualitatively speaking, this factor arises due to the
degeneracy pressure of filled fermionic levels that disfavor
scattering of neutrinos to the direction $\theta$ when $\rho^{\theta}$
is sufficiently close to 1.

The general form of the collision rate [Eq.~\eqref{coll:nuc}] is too
complex to consider in the linearized stability analysis we pursue
here. In order to make the problem tractable and to include the effect
of angle dependent scattering in a basic way, we approximate the
collision integral in two ways. First, we ignore the effect of Pauli
blocking in the second term of Eq.~\eqref{coll:nuc}.  This will be
problematic in regimes of high neutrino degeneracy but is not too
drastic an approximation during the epoch of
weak-decoupling/nucleosynthesis that we're ultimately interested in
understanding. As a further simplifying approximation, we consider
only the neutrino-nucleon-like scattering processes  $\nu_\alpha N \to
\nu_\alpha N$, which is flavor $\alpha$ independent, neutral current
scattering.  Incorporating the aforementioned assumptions allows the
reduction of the general form of scattering matrices given in
Ref.~\cite{Blaschke:2016xxt}. 
Using Eqs.~(3.21)-(3.22)   (Eqs. (60)-(61))  
in the e-print archive (published) version of Ref.~\cite{Blaschke:2016xxt}, the 
form of the collision integral can be calculated by explicitly taking the inner 
product in 2+1 dimensional space. 
We find that the matrix-valued
gain and loss collision contributions are proportional to the
flavor-space identity matrix
\begin{align}
\label{coll:form}
\Pi^{\textrm{loss/gain}} 
& \sim \Big[ c_{1} - c_{2} \cos(\theta-\theta^{\prime}) \Big] 
\times \mathds{1}_{2\times2}
\end{align}
where, $c_{1}$ and $c_{2}$ are constants. In three dimensional space
one has  $c_{1}  \propto G_{F}^{2}n_{\textrm{nuc}}p^{2}$ with
proportionality  constant of order unity. For equal number of protons
and neutrons in the early Universe the ratio $c_{1}/c_{2}$ is fixed
and is equal   to 4.83, a value which we use for rest of the paper.

We note that the above form for the approximate collision rate in
Eq.~\eqref{coll:form} is energy-local; that is, there is no energy
transport. And it is angle dependent, being a function of the angle of
scattering $\theta'$ relative to the incoming neutrino angle $\theta$.

Combining Eqs.~\eqref{eom} and \eqref{coll:form} we obtain the
complete equation describing the evolution of flavor density matrices
in presence of collective effects and collisions.

\section{Linear stability analysis}
\label{linearstability}

Nonlinear neutrino self-interaction in the equations of motion for the
density matrix [Eqs.~\eqref{eqn:dm_eom}] presents a computationally
challenging scenario for the general case, particularly when $\vec{k}\ne 0$.
In the presence of seed spatial inhomogeneity perturbations, 
the reduction of the number of
independent variables by symmetry is no longer possible.  It is,
however, possible to gain insight into the stability properties
of Eqs.~\eqref{eqn:dm_eom}  through a linear stability
analysis~\cite{Banerjee:2011fj}. In this section we describe the
formalism and later,  in Sec.~\ref{sec:concl}, we
discuss the reliability and limitations of linear stability analyses
in the context of collective neutrino oscillations.

The linear stability analysis of collective neutrino oscillations is
predicated on the basis that, for significant neutrino flavor
oscillations to occur, the off-diagonal elements of the Hamiltonian
should be comparable in magnitude to the diagonal elements. Since
$\mathsf{H}_{0}$ is diagonal (neglecting vacuum mixing),
this can occur if the off-diagonal component of density matrices
increase rapidly with time. 

In order to investigate the conditions (parameter ranges) under which
exponential growth of the Fourier modes may occur, we assume that the
density matrices have a form describing the presence of seed
inhomogeneity perturbations ($\Delta^\theta(t,\vec{x})$  and  $\epsilon^{\theta}(t,\vec{x})$  for neutrinos 
and $\bar \Delta^\theta(t,\vec{x})$  and  $\bar \epsilon^{\theta}(t,\vec{x})$  for antineutrinos):
\begin{align}
& \rho^{\theta}(t,\vec{x}) =
\begin{pmatrix}
1-\Delta^\theta(t,\vec{x}) & \epsilon^{\theta}(t,\vec{x}) \\
 \epsilon^{\theta*}(t,\vec{x}) & \Delta^\theta(t,\vec{x})
\end{pmatrix}   
& |\epsilon^{\theta}(t,\vec{x})|,|\Delta^\theta(t,\vec{x})| \ll 1
\nonumber \\ 
&\bar{\rho}^{\theta}(t,\vec{x}) =
\begin{pmatrix}
1-\bar{\Delta}^\theta (t,\vec{x}) & \bar{\epsilon}^{\theta}(t,\vec{x}) \\
\bar{\epsilon}^{\theta*}(t,\vec{x}) & \bar{\Delta}^\theta (t,\vec{x})
\end{pmatrix} 
& |\bar{\epsilon}^{\theta}(t,\vec{x})|,|\bar{\Delta}^\theta(t,\vec{x})| \ll 1.
\label{rho:lin}
\end{align}
We leave the origin of these seed inhomogeneity perturbations
unspecified. They will be present in generic theories of the early
universe unless expressly forbidden by some exact symmetry. 

The perturbations in the diagonal terms, $\Delta^\theta(t,\vec{x})$ and
$\bar{\Delta}^\theta (t,\vec{x})$ can be ignored upon linearization of the
equations of motion, because they contribute only at second order.  In terms of
the Fourier modes, Eq.~\eqref{rho:lin} reads  
\begin{eqnarray}
\rho^{\theta}_{\vec{k}}(t)= 
\begin{pmatrix}
\delta_{\vec{k},0} & \epsilon^{\theta}_{\vec{k}}(t) \\
 \epsilon^{\theta*}_{-\vec{k}}(t) & 0 
\end{pmatrix} & 
\bar{\rho}^{\theta}_{\vec{k}}(t)= 
\begin{pmatrix}
\delta_{\vec{k},0} & \bar{\epsilon}^{\theta}_{\vec{k}}(t) \\
\bar{\epsilon}^{\theta*}_{-\vec{k}}(t) & 0 
\end{pmatrix}~.
\end{eqnarray}
The equations for $\epsilon_{\vec{k}}$ for the various Fourier modes
$\vec{k}$ are decoupled at linear order in the perturbations
$\eps^\theta_{\vec{k}}$, $\bar\eps^\theta_{\vec{k}}$:
\begin{align}
&i \frac{\partial}{\partial t} \epsilon^{\theta}_{\vec{k}}(t)
= \vec{v}\cdot \vec{k} \epsilon^{\theta}_{\vec{k}}(t) 
- \omega \epsilon^{\theta}_{\vec{k}}(t) \nonumber \\
&+ 2 \pi \mu (1-\alpha) \epsilon^{\theta}_{k}
- \mu \int_{0}^{2\pi} d\theta^{\prime} 
(\epsilon^{\theta^{\prime}}_{k}-\alpha 
\bar{\epsilon}^{\theta^{\prime}}_{k}) 
(1-\cos(\theta-\theta^{\prime})) \nonumber \\
&-i2\pi c_{1}\epsilon^{\theta}_{\vec{k}}(t) 
+ i\int_{0}^{2\pi}   d\theta^{\prime}
(c_{1}-c_{2}\cos(\theta-\theta^{\prime})) 
\epsilon^{\theta^{\prime}}_{\vec{k}}(t),\nonumber \\
\label{off:eom}
&i \frac{\partial}{\partial t} \bar{\epsilon}^{\theta}_{\vec{k}}(t) 
= \vec{v}\cdot \vec{k} \bar{\epsilon}^{\theta}_{\vec{k}}(t) 
+ \omega \bar{\epsilon}^{\theta}_{\vec{k}}(t) \nonumber \\
&+ 2 \pi \mu (1-\alpha) \bar{\epsilon}^{\theta}_{k}
- \mu \int_{0}^{2\pi} d\theta^{\prime} 
(\epsilon^{\theta^{\prime}}_{k}-\alpha 
\bar{\epsilon}^{\theta^{\prime}}_{k}) 
(1-\cos(\theta-\theta^{\prime}))\nonumber \\
&-i2\pi c_{1}\bar{\epsilon}^{\theta}_{\vec{k}}(t) 
+ i\int_{0}^{2\pi}  d\theta^{\prime}   
(c_{1}-c_{2}\cos(\theta-\theta^{\prime})) 
\bar{\epsilon}^{\theta^{\prime}}_{\vec{k}}(t) .
\end{align}
For a given set of parameters, exponential growth of $\epsilon$ and
$\bar{\epsilon}$ implies flavor instability. 
The dependence of collective neutrino oscillations on hierarchy can be 
viewed as being a result of the relative sign between the second term 
on right-hand side of Eq.~\eqref{off:eom}, which 
is proportional to the vacuum term, and the third and forth terms, which 
is proportional to the self-interaction strength $\mu$. However, it should 
be noted that the relative sign between the vacuum and self-interaction 
term can also be changed by going from $\alpha<1$ to $\alpha>1$. 
All the results therefore for $1-\alpha>0$ with normal hierarchy 
are identical to those for $1-\alpha<0$ with inverted hierarchy.
There are two different
ways in which we can investigate the flavor instability of the system.
One is the brute-force method in which we discretize the angular
dependence of $\epsilon$ and $\bar{\epsilon}$ and solve the resulting 
system of coupled differential equations.  A major drawback of
this approach is that sometimes we get `spurious' flavor instabilities
as a result of discretization which reduce as we increase the number
of angle-bins\cite{Abbar:2015mca}. 

The second method is a bit more formal but does not have the problem
of spurious instabilities. Based on the fact that we are investigating
collective modes, we seek solutions to Eqs.~(\ref{off:eom}) of the form
\begin{align}
&\epsilon^{\theta}_{\vec{k}}(t) = Q^{\theta}_{\vec{k}} \exp(-i\Omega_{\vec{k}} t) \nonumber\\
&\bar{\epsilon}^{\theta}_{\vec{k}}(t) =
\overline{Q}^{\theta}_{\vec{k}} \exp(-i\Omega_{\vec{k}} t).
\label{Q:form}
\end{align}
Existence of a solution with $\kappa > 0$, where
\begin{align}
\label{eqn:def_kappa}
\kappa\equiv\text{Im }(\Omega_{\vec{k}}),
\end{align}
implies the existence of flavor instabilities.  Combing
Eq.~\eqref{off:eom} and Eq.~\eqref{Q:form} we obtain
\begin{align}
\label{Q:wcoll}
&\left[\Omega_{\vec{k}} - \vec{v}\cdot \vec{k} + \omega -2\pi\mu(1-\alpha) 
       + i2\pi c_{1} \right]Q^{\theta}_{\vec{k}} \nonumber \\ 
&= -\mu \int_{0}^{2\pi} d\theta^{\prime} 
(Q^{\theta^{\prime}}_{\vec{k}}-
  \alpha \overline{Q}^{\theta^{\prime}}_{\vec{k}}) 
  (1-\cos(\theta-\theta^{\prime})) \nonumber \\
& +i \int_{0}^{2\pi}d\theta^{\prime}
  Q^{\theta^{\prime}}_{\vec{k}}
  (c_{1}-c_{2}\cos(\theta-\theta^{\prime})),
  \end{align}
for the neutrino modes, and
\begin{align}
\label{barQ:wcoll}
&\left[\Omega_{\vec{k}} - \vec{v}\cdot \vec{k} - \omega -2\pi\mu(1-\alpha) 
        + i2\pi c_{1} \right]\overline{Q}^{\theta}_{\vec{k}} \nonumber \\
&= - \mu \int_{0}^{2\pi} d\theta^{\prime} 
(Q^{\theta^{\prime}}_{\vec{k}}
-\alpha \overline{Q}^{\theta^{\prime}}_{\vec{k}}) 
(1-\cos(\theta-\theta^{\prime})) \nonumber \\
&+i \int_{0}^{2\pi}d\theta^{\prime}
\overline{Q}^{\theta^{\prime}}_{\vec{k}}
(c_{1}-c_{2}\cos(\theta-\theta^{\prime})),
\end{align}
for the antineutrino modes.

\begin{figure*}[!ht]
\includegraphics[width=0.49\textwidth]{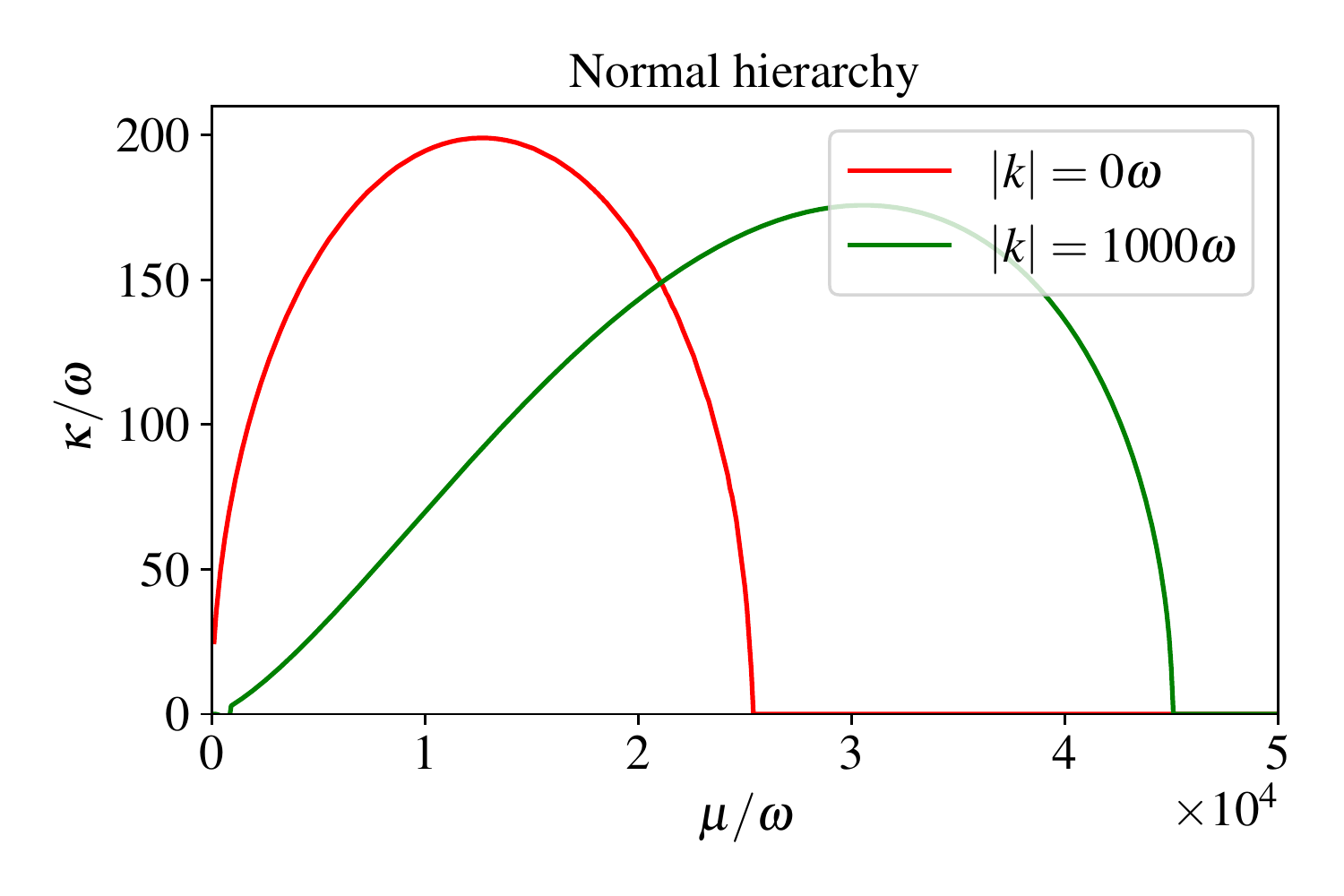}
\includegraphics[width=0.49\textwidth]{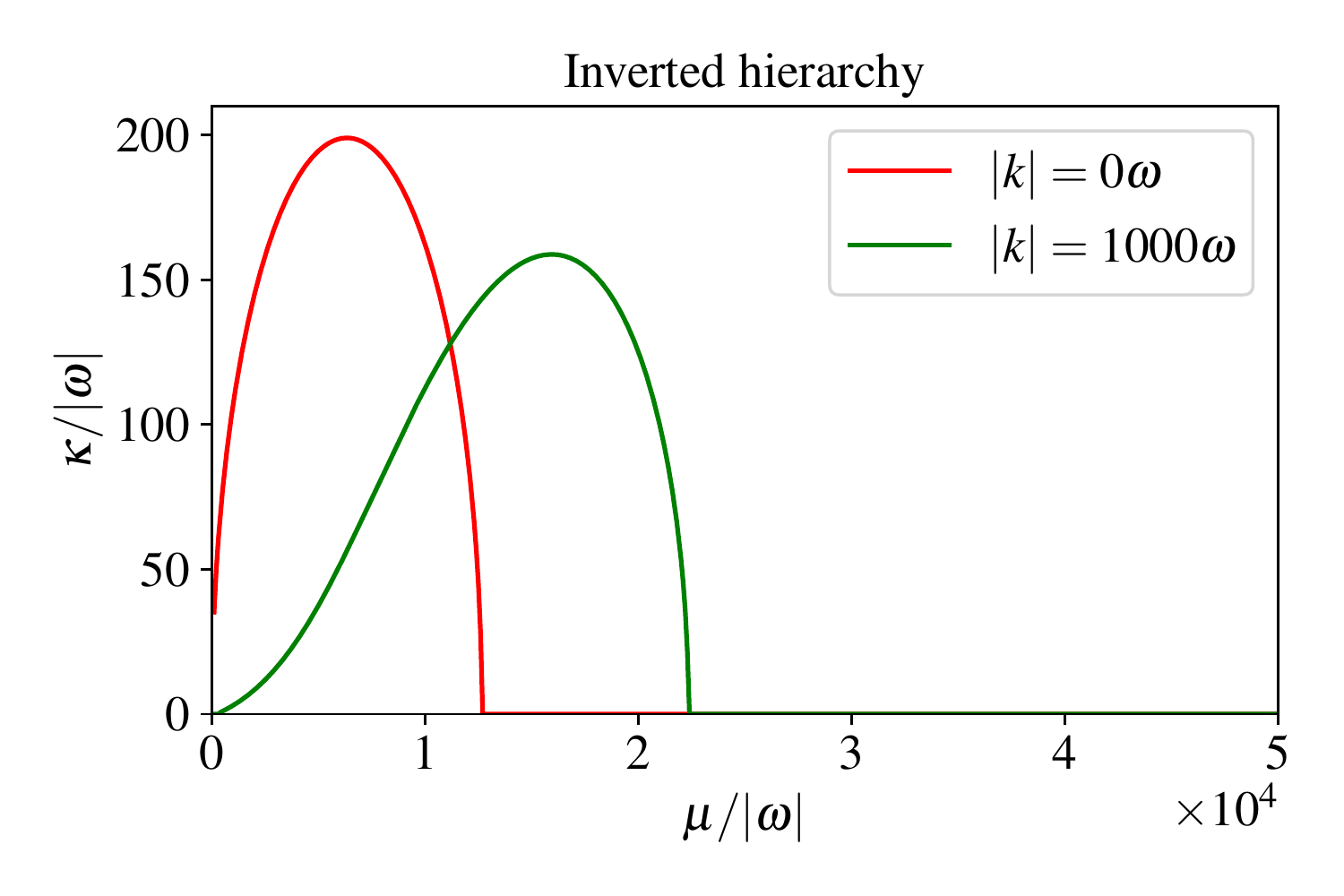}
\caption{\label{nocoll}Imaginary parts, $\kappa$ of the unstable mode
frequency $\Omega_{\vec{k}}$ as a function of the ratio of the
strength of the effective neutrino coupling $\mu$ [Eq.~\eqref{eqn:Hnn}]
to that of the scaled vacuum frequency $\omega$ for the collisionless
system. The normal hierarchy is displayed in the left panel and the
inverted in the right panel.  The red curve is the homogeneous mode
$\vec{k}=0$ and the green curve is a high wave number, small scale
inhomogeneity with $|\vec{k}|=1000 \omega$ and $\alpha=0.99$.}
\end{figure*}

Recognizing that the solutions to the above  equations for the collective mode 
amplitudes $Q^{\theta}_{\vec{k}}$ and $\overline{Q}^{\theta}_{\vec{k}}$ are
spanned by the functions $\{1, \cos\theta, \sin\theta\}$  
leads to the  ansatz 
\begin{align}
\label{eqn:Q}
Q^{\theta}_{\vec{k}} &= -\frac{\mu}{D_{\vec{k}}}
\left[(a_{\vec{k}}-\alpha\bar{a}_{\vec{k}})
-\sin\theta(s_{\vec{k}}-\alpha\bar{s}_{\vec{k}})
-\cos\theta(c_{\vec{k}}-\alpha\bar{c}_{\vec{k}})\right]\nonumber\\ 
& + \frac{i}{D_{\vec{k}}}
\left[c_{1}a_{\vec{k}}-c_{2}\sin\theta s_{\vec{k}} 
-c_{2}\cos\theta c_{\vec{k}}\right]\\
\label{eqn:barQ}
\overline{Q}^{\theta}_{\vec{k}} &= -\frac{\mu}{\overline{D}_{\vec{k}}}
\left[(a_{\vec{k}}-\alpha\bar{a}_{\vec{k}})
-\sin\theta(s_{\vec{k}}-\alpha\bar{s}_{\vec{k}})
-\cos\theta(c_{\vec{k}}-\alpha\bar{c}_{\vec{k}})\right]
\nonumber\\ 
& + \frac{i}{\overline{D}_{\vec{k}}}
\left[c_{1}\bar{a}_{\vec{k}}-c_{2}\sin\theta \bar{s}_{\vec{k}} 
-c_{2}\cos\theta \bar{c}_{\vec{k}}\right],
\end{align}
where we have defined the collective mode functions 
$a_{\vec{k}}$, corresponding to modes independent of the angle
$\theta$ and the angular-dependent modes
$s_{\vec{k}}$ and $c_{\vec{k}}$ (and their barred counterparts).  
We have abbreviated the denominators as:
\begin{align}
\label{eqn:DbarD}
D_{\vec{k}} &= D(\omega,\Omega_{\vec{k}}) 
= {\left[\Omega_{\vec{k}} - \vec{v}\cdot \vec{k} 
+ \omega -2\pi\mu(1-\alpha)\right] + i2\pi c_{1}}, \nonumber \\
\overline{D}_{\vec{k}} &= D(-\omega,\Omega_{\vec{k}}).
\end{align}

The collective mode functions are projected from Eqs.~\eqref{eqn:Q} and \eqref{eqn:barQ} as
\begin{align}
&a_{\vec{k}} = \int_{0}^{2\pi} d\theta\, Q^{\theta}_{\vec{k}} &  
&\bar{a}_{\vec{k}} = \int_{0}^{2\pi} d\theta\, \overline{Q}^{\theta}_{\vec{k}} \cr
&s_{\vec{k}} = \int_{0}^{2\pi} d\theta \sin\theta Q^{\theta}_{\vec{k}} &  
&\bar{s}_{\vec{k}} = \int_{0}^{2\pi} d\theta \sin\theta \overline{Q}^{\theta}_{\vec{k}} \cr
&c_{\vec{k}} = \int_{0}^{2\pi} d\theta \cos\theta Q^{\theta}_{\vec{k}} &  
&\bar{c}_{\vec{k}} = \int_{0}^{2\pi} d\theta \cos\theta \overline{Q}^{\theta}_{\vec{k}} 
\label{asc:newdef}
\end{align}    
Consistency among these relations yields linear equations for the
collective modes, given by:

\begin{widetext}
\begin{align}
\begin{pmatrix}
-J_{1}[1](c_{1})-1         & \alpha I_{1}[1]           & J_{1}[\sin\theta](c_{2})           & -\alpha I_{1}[\sin\theta]           & J_{1}[\cos\theta](c_{2})           & -\alpha I_{1}[\cos\theta]            \\
-I_{2}[1]                  & -J_{2}[1](c_{1})-1        & I_{2}[\sin\theta]                  & J_{2}[\sin\theta](c_{2})            & I_{2}[\cos\theta]                  & J_{2}[\cos\theta](c_{2})             \\
-J_{1}[\sin\theta](c_{1})  & \alpha I_{1}[\sin\theta]  & J_{1}[\sin^{2}\theta](c_{2})-1     & -\alpha I_{1}[\sin^{2}\theta]       & J_{1}[\sin\theta\cos\theta](c_{2}) & -\alpha I_{1} [\sin\theta\cos\theta] \\
 -I_{2}[\sin\theta]        & -J_{2}[\sin\theta](c_{1}) & I_{2}[\sin^{2}\theta]              & J_{2}[\sin^{2}\theta](c_{2})-1      & I_{2}[\sin\theta\cos\theta]        & J_{2}[\sin\theta\cos\theta](c_{2})   \\
 -J_{1}[\cos\theta](c_{1}) & \alpha I_{1}[\cos\theta]  & J_{1}[\sin\theta\cos\theta](c_{2}) & -\alpha I_{1}[\sin\theta\cos\theta] & J_{1}[\cos^{2}\theta](c_{2})-1     & -\alpha I_{1}[\cos^{2}\theta]        \\
 -I_{2}[\cos\theta]        & -J_{2}[\cos\theta](c_{1}) & I_{2}[\sin\theta\cos\theta]        & J_{2}[\sin\theta\cos\theta](c_{2})  & I_{2}[\cos^{2}\theta]              & J_{2}[\cos^{2}\theta](c_{2})-1
\end{pmatrix}
\begin{pmatrix}
a_{\vec{k}} \\ \bar{a}_{\vec{k}} \\ s_{\vec{k}} \\ \bar{s}_{\vec{k}} \\ {c}_{\vec{k}} \\ \bar{c}_{\vec{k}}
\end{pmatrix}
&=0,
\label{uglymatrix}
\end{align}
\end{widetext}
\noindent in terms of the functionals 
\begin{align}
I_{1}[f] &= \int_{0}^{2\pi}d\theta\,
  \frac{\mu f}{D_{\vec{k}}}, &
I_{2}[f] &= \int_{0}^{2\pi}d\theta\,
  \frac{\mu f}{\overline{D}_{\vec{k}}},\\
I_{3}[f] &= \int_{0}^{2\pi}d\theta\,
  \frac{if}{D_{\vec{k}}}, &
I_{4}[f] &= \int_{0}^{2\pi}d\theta\,
  \frac{if}{\overline{D}_{\vec{k}}},
\end{align}
with $J_{1,2}$ defined as
\begin{align}
J_{1}[f](c) &= I_{1}[f]-c I_{3}[f], \nonumber \\
J_{2}[f](c) &= -\alpha I_{2}[f] - c I_{4}[f]~. 
\end{align}

\begin{figure*}
\includegraphics[width=0.49\textwidth]{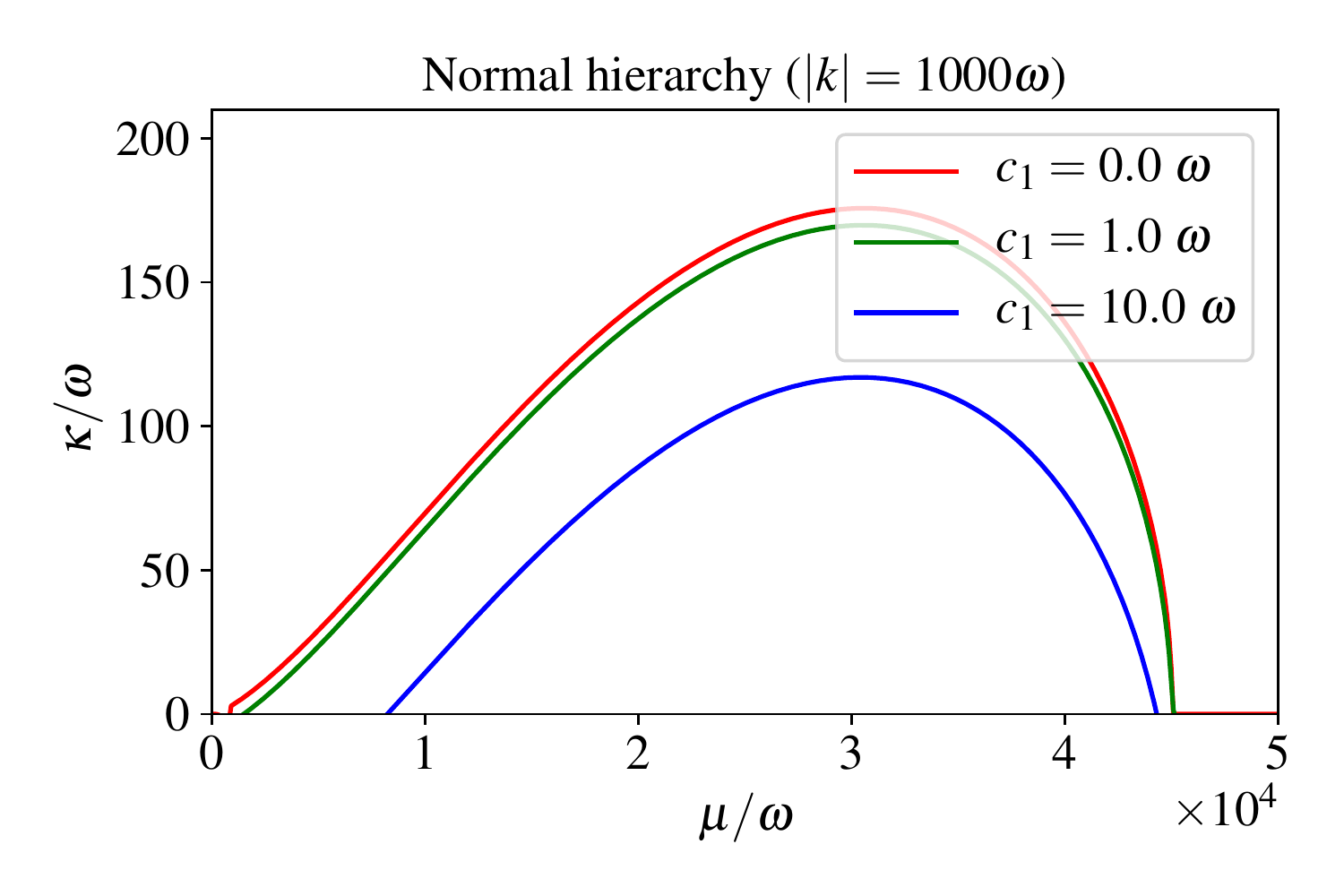}
\includegraphics[width=0.49\textwidth]{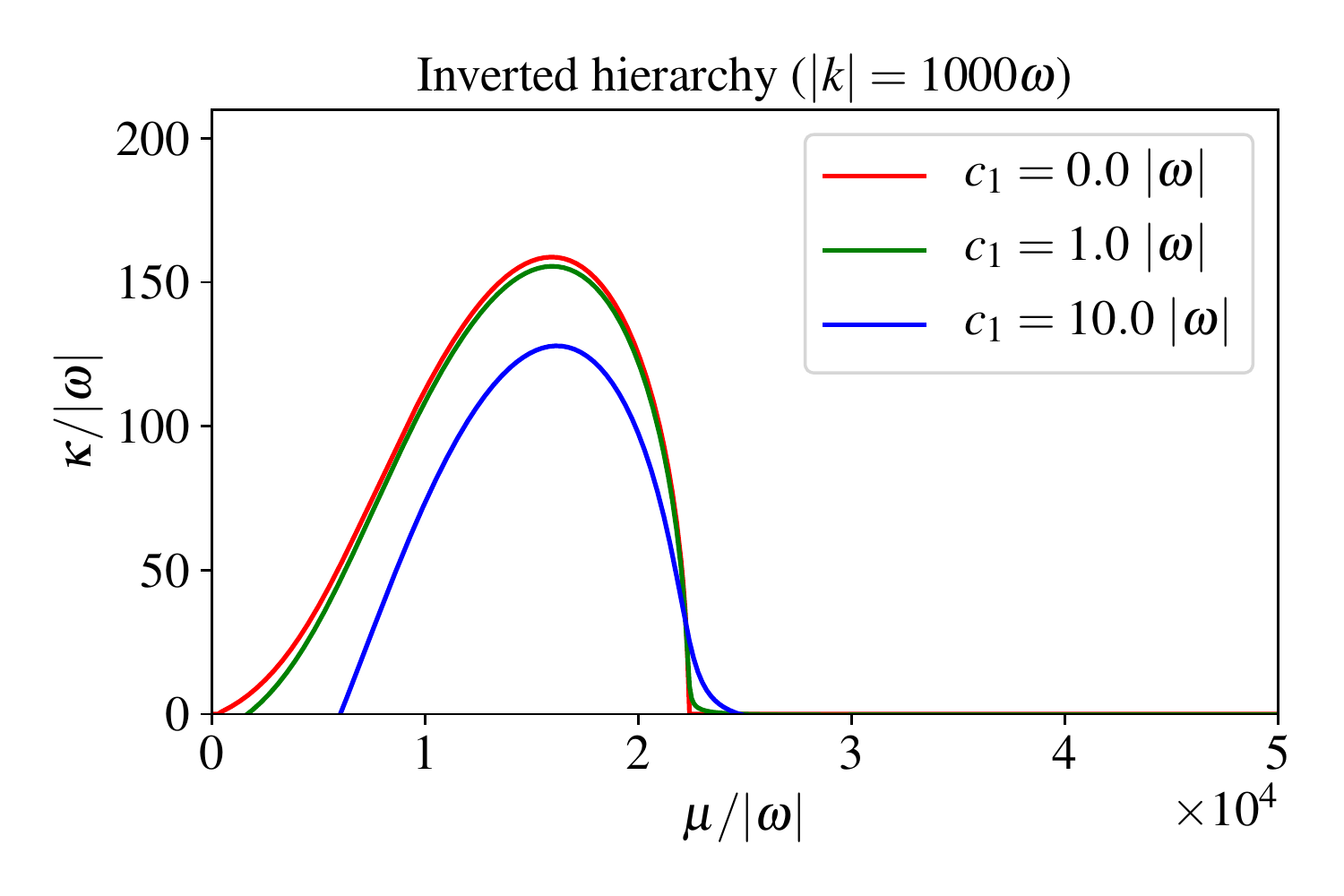}
\caption{\label{fig:k1e4}The quantity $\kappa/\omega$ of small-scale
$|\vec{k}|=1000 \omega $ inhomogeneity as a function of
$\mu/\omega$ [Eq.~\eqref{eqn:Hnn}] for various values of the strength
of the incoherent collision term, Eq.~\eqref{coll:form} as measured by
the constant $c_1$. The normal hierarchy is displayed in the left
panel and the inverted in the right panel. In each panel, the red
curve is the collisionless case ($c_1 = 0$), the green curve is the
intermediate value of collision strength ($c_1= 1.0\, \omega$), and
blue is strong coupling ($c_1 = 10.0 \, \omega$). The ratio 
$\bar{\nu}_{e}/\nu_{e}$ defined as $\alpha$ is fixed at 0.99.
}
\end{figure*}

We determine the values of $\Omega_{\vec{k}}$ for which
Eq.~\eqref{uglymatrix} is satisfied by requiring the determinant of the
$6\times 6$ matrix vanishes. As the determinant of the matrix is a
$6^{\textrm{th}}$-order polynomial we obtain up to six different
values of $\Omega_{\vec{k}}$ for each value of $\vec{k}$. While
multiple solutions are possible, the instability is dominated by the
$\Omega_{\vec{k}}$ with largest positive imaginary part. We have
numerically solved for $\Omega_{\vec{k}}$ by using standard root
finding numerical algorithms. For the results presented in the next
section we use the open source GNU Scientific Library
(GSL)~\cite{Gough:2009:GSL:1538674}.

Note that for $\vec{k}=0$ (homogeneous) modes  the matrix in
Eq.~\eqref{uglymatrix} has a block diagonal form with three $2\times 2$
blocks corresponding to subspaces spanned by the pairs of mode
functions $\{a_{\vec{k}}$, $\bar{a}_{\vec{k}}\}$, $\{s_{\vec{k}}$,
$\bar{s}_{\vec{k}}\}$, and $\{c_{\vec{k}}$, $\bar{c}_{\vec{k}}\}$;
these subspaces are decoupled.  The latter two subspaces are identical
(those for the $s_{\vec{k}}$, $\bar{s}_{\vec{k}}$ and $c_{\vec{k}}$,
$\bar{c}_{\vec{k}}$ blocks).  In this case, the determinant can be
calculated analytically and the solutions for
$\Omega_0=\Omega_{\vec{k}=0}$ are given by
\begin{equation}
\label{eq:Oma}
\Omega_{0,\pm}^{(a)} = \pi \mu (1 - \alpha) 
\pm \sqrt{ \omega^2  + 2 \pi  \mu (1 + \alpha) \omega 
   + \pi^2 \mu^2 (1 - \alpha)^2 },
\end{equation}
for  the $a_{\vec{k}=0}=a_0,\bar{a}_{\vec{k}=0}=\bar{a}_0$ block, and
\begin{align}
\label{eq:Omsc}
\Omega_{0,\pm}^{(s,c)} &= \frac{5}{2}  
   \pi \mu (1 - \alpha) - i \pi (2 c_1 + c_2) \nonumber \\ 
&\pm   \sqrt{ \omega^2  -   \pi  \mu (1 + \alpha) \omega 
   + \frac{1}{4}   \pi^2 \mu^2 (1 - \alpha)^2},
\end{align}
for the $s_{0}$, $\bar{s}_{0}$ and $c_{0}$, $\bar{c}_{0}$ blocks. 
Despite the fact that the $\vec{k}=0$ modes are homogeneous, it is
instructive to study these results for this $\vec{k}=0$ mode since
similar behavior is observed for the inhomogeneous $\vec{k}\ne 0$
modes.

For the inverted hierarchy ($\omega<0$),  in absence of collisions the angle-dependent
collective frequencies $\Omega_{0,\pm}^{(s,c)}$ are real and therefore
do not give rise to exponential growth. 
On the other hand, Eqs.\eqref{eq:Oma} indicate that for the inverted
hierarchy ($\omega < 0$), the argument of the square-root
$\kappa^2=\omega^2-2\pi\mu(1+\alpha)|\omega|+\pi^2\mu^2(1-\alpha)^2$
becomes negative for values
\begin{align}
\label{eqn:mu_unstable_limits}
\frac{1}{4 \pi} \lesssim \frac{\mu}{|\omega|} \lesssim  \frac{4}{\pi\delta^2},
\end{align}
where we have  defined $\alpha = 1 - \delta$ and expanded for $\delta \ll 1$ 
(in our numerical studies we  take $\delta = 10^{-2}$). 
Figure \ref{nocoll} shows $\kappa/|\omega|$ plotted against the neutrino effective
self-coupling $\mu/|\omega|$; the left panel shows two $\vec{k}$ modes (red and
green curves) for the normal (left panel) and inverted (right panel)
hierarchies.  The effective coupling inequalities for the inverted
hierarchy described by Eqs.~\eqref{eqn:mu_unstable_limits} are clearly
exhibited in the right panel of Fig.~\ref{nocoll} by the curve
corresponding to $\vec{k}=0$ (the red curve).  The left-most
inequality in Eqs.~\eqref{eqn:mu_unstable_limits} corresponds to the
region $\mu/|\omega| \approx 0$;   the right-most inequality corresponds to where
the $|\vec{k}|=0$ intersects the abscissa near $\mu/|\omega| \simeq 1.27 \times 10^{4}$.

Equation \eqref{eq:Oma} demonstrates that for inverted hierarchy ($\omega < 0$) only 
$\theta$-independent instabilities can develop (proportional to $a_{\vec{k}}$ and $\bar
a_{\vec{k}}$). Moreover, and perhaps not too  surprisingly, the
direction-changing form for the elastic collision rate we employ
[Eq.~\eqref{coll:form}] in this exploratory calculation has no effect
on such instabilities: the strength of the collisions, as measured by
the quantities $c_{1,2}$, do not feature in the expression for
$\Omega_{0,\pm}^{(a)}$. This is potentially an important observation
since, if this behavior obtains for the full, unapproximated collision
rate then nonlinear mode coupling effects, present in the full QKEs,
could propagate the instability to other, inhomogeneous modes. This
effect could result in a continuous ``feeding'' of inhomogeneity that
may persist for times long compared to other dynamical time scales,
despite collisional damping in the inhomogeneous modes.

For the normal hierarchy ($\omega>0$), the angle-independent
collective frequencies $\Omega_{0,\pm}^{(a)}$ are real and therefore
do not give rise to exponential growth.  On the other hand,
$\theta$-dependent instabilities (proportional to $s_0,c_0$ and $\bar
s_0, \bar c_0$) described by Eq.~\eqref{eq:Omsc} may develop.  
In this case, the growth rate due to the
imaginary part of $\Omega^{(s,c)}_{0,\pm}$ is indeed damped by the
direction-changing collisions, as shown by the term proportional to $2
c_1 + c_2$ in Eq.~\eqref{eq:Omsc}. The numerical results discussed in
the next section match the analytic results at $\vec{k}=0$ and show
that a similar behavior persists for $\vec{k} \neq 0$: for normal
hierarchy $\theta$-dependent instabilities ($s_{\vec{k}}$, $\bar
s_{\vec{k}}$ and  $c_{\vec{k}}$, $\bar c_{\vec{k}}$) can develop,
while for the inverted hierarchy only the $\theta$-independent modes
$a_{\vec{k}}$ and $\bar a_{\vec{k}}$ are subject to unstable,
exponential growth.

\begin{figure*}
\includegraphics[width=0.49\textwidth]{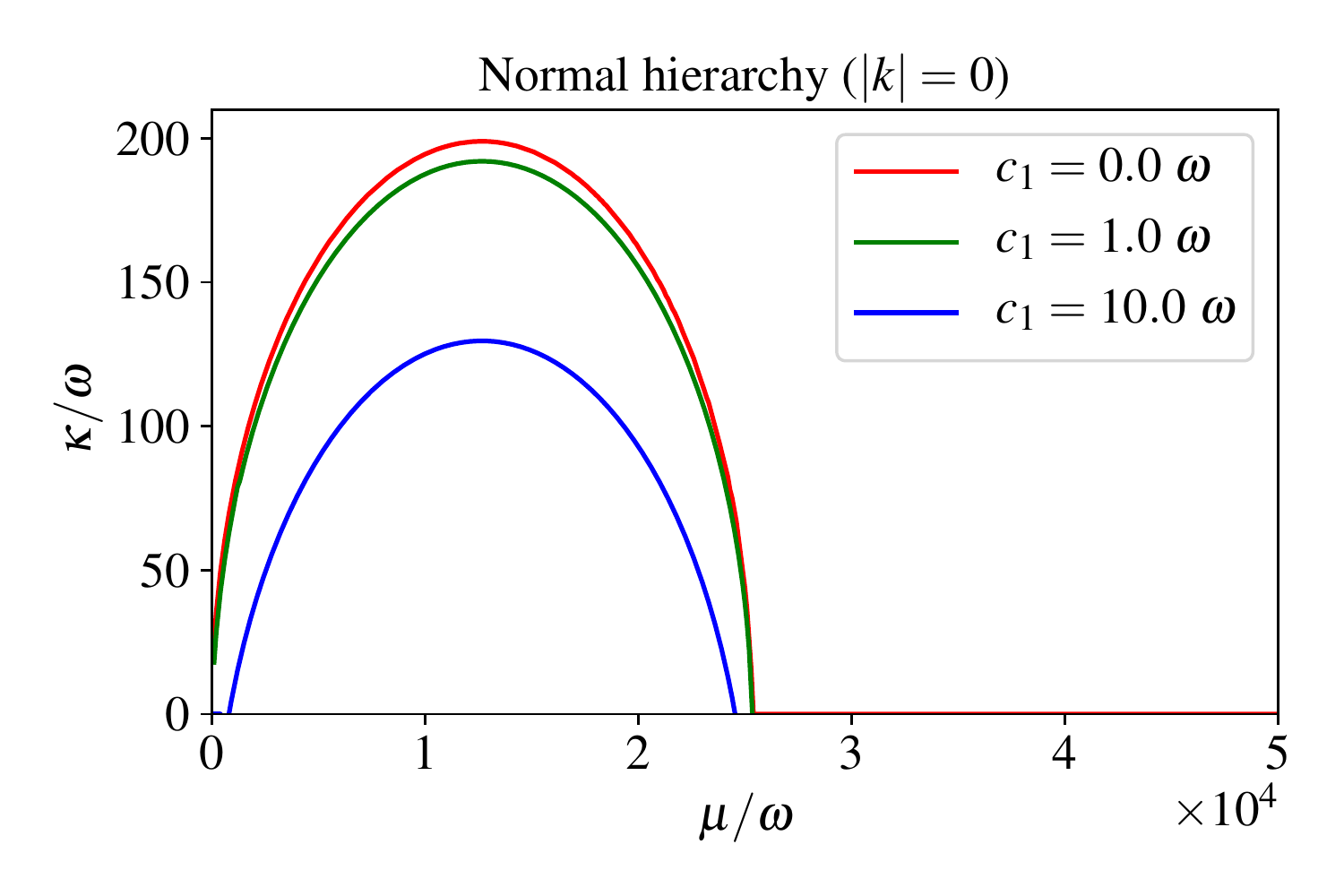}
\includegraphics[width=0.49\textwidth]{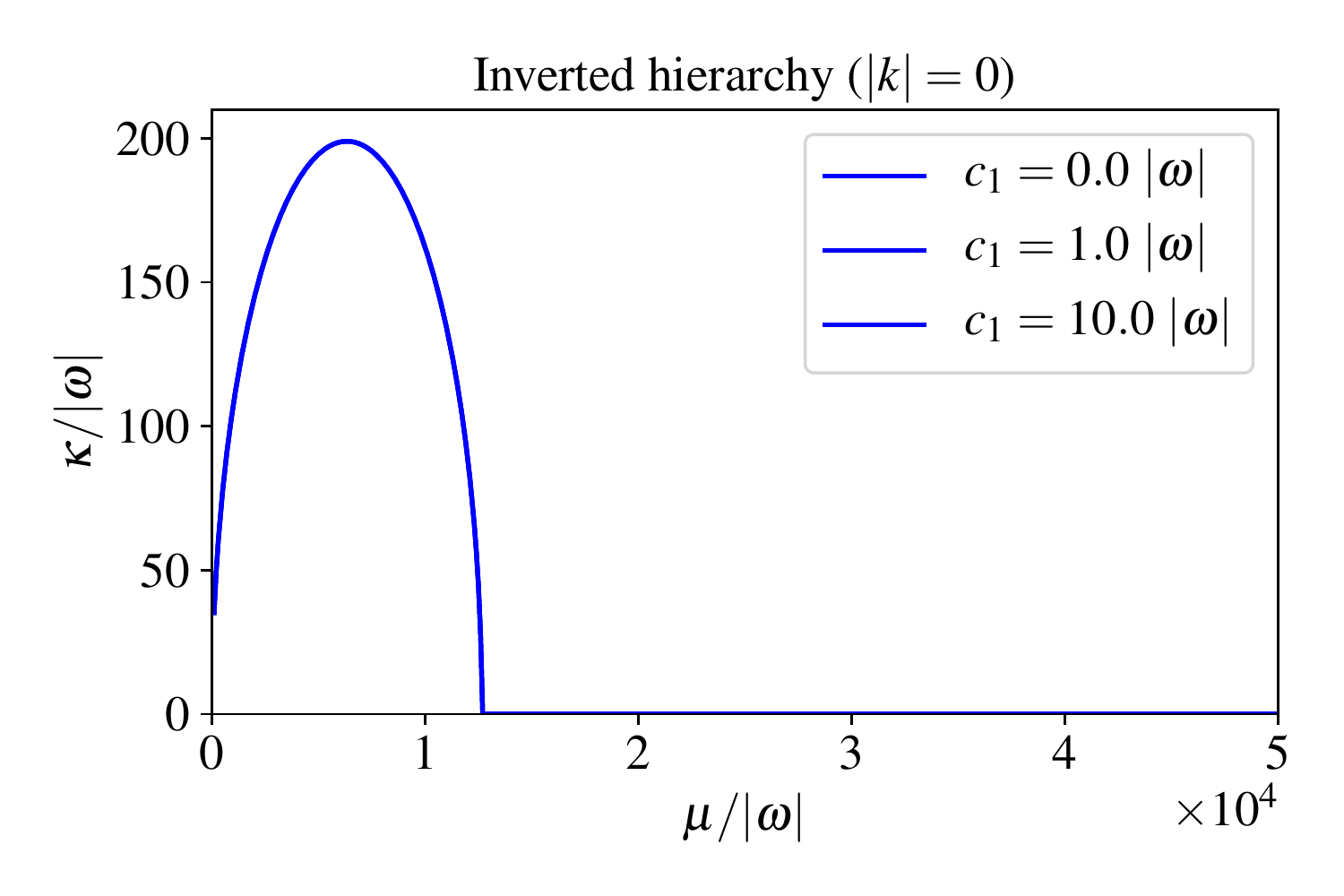}
\caption{\label{fig:k0}The quantity $\kappa/\omega$ for large-scale
$|\vec{k}|\approx 0$ inhomogeneity as a function of
$\mu/\omega$ [Eq.~\eqref{eqn:Hnn}] for various values of the strength
of the incoherent collision term, Eq.~\eqref{coll:form} as measured by
the constant $c_1$; see Fig.~\ref{fig:k1e4} for the description.}
\end{figure*}

\section{Results}
\label{sec:results}

In this section we present numerical results for the stability analysis of 
collective modes with $\vec{k}\ne 0$ corresponding to inhomogeneous
flavor distributions in configuration space. 
We continue to take the ratio of electron anti-neutrinos to electron
neutrinos, $\alpha = 0.99$, as in the previous section. As shown in
Fig.~\ref{fig:k1e4}, for the collisionless case ($c_1=0=c_2$), the
inhomogeneous modes with $|\vec{k}|=1000 \omega$ 
(we use $n=10^{4}$ and $2\pi/L=0.1 \omega$) 
are unstable in both the normal (left panel of the figure) and inverted
(right panel) mass hierarchies. The effective neutrino self-coupling,
$\mu$ over which instabilities exist, we note, is different for normal
and inverted hierarchies.  However, for both mass hierarchies the
small scale inhomogeneous modes are unstable over a larger range of
$\mu$ than for the homogeneous modes.

In the presence of collisions we would generally expect a suppression
of instabilities for sufficiently strong interaction rates. This
expectation is driven principally by the observation that incoherent
scattering of the form of Eq.~\eqref{coll:form}, due to its angular
dependence, will tend to redistribute regions of neutrino density
toward the mean.  The expectation of the suppression is supported by
the form of Eq.~\eqref{eq:Omsc}, specifically the term $-i \pi (2c_1 +
c_2)$.  Since $c_{1,2} >0$,   this term contributes an
exponentially damped factor to the collective mode frequency
$\Omega^{(s,c)}_{0,\pm}$.

Indeed, for large values of $k \sim 1000 \omega$ we find suppression of
instabilities for both mass hierarchies, as shown in
Fig.~\ref{fig:k1e4}. As in Fig.~\ref{nocoll}, the left and right panels
give results for the values of $\kappa/\omega$ for the normal and
inverted hierarchies, respectively. In each hierarchy, the effect of
varying the strength of the collision term, measured by the magnitude
of $c_1>0$, is shown by the various curves, as detailed in the figure
caption.  In each panel, we observe a reduction in the width of the
region of instability with increasing $c_1$.

However, as previously indicated by Eq.~\eqref{eq:Oma}, for smaller
values of $k$ ($k \sim $ few $\times 100\omega$) including zero, we find a suppression
due to collisions  only in the case of normal mass hierarchy,
not in the inverted mass hierarchy. Results for the $\vec{k}=0$
homogeneous modes have been obtained both analytically, through
Eq.~\eqref{eq:Oma} and numerically via diagonalization of the $6\times
6$ matrix appearing in Eq.~\eqref{uglymatrix}.  Figure \ref{fig:k0} shows the
numerical results for the homogeneous mode for various values of the
incoherent scattering strength parameter $c_1$; it agrees with
Eq.~\eqref{eq:Oma} at high precision for all values of $c_1$ considered
in this work. 

\section{Conclusion}
\label{sec:concl}


We have studied the stability of a two-dimensional dense neutrino gas with respect to spatial inhomogeneity. 
In the absence of incoherent collisional effects  we have   found that  the  system exhibits growth of seed inhomogeneity 
due to nonlinear coherent neutrino self-interactions and that the effect depends on the  neutrino mass spectrum.  
For  normal hierarchy   spatial instability  exists over a larger range
of neutrino number density compared to the inverted case.

We have further considered  the effect of  elastic  (i.e. energy conserving) incoherent (direction-changing)  
collisions of the neutrinos with a  static background of heavy, nucleon-like scatterers,  
in order to validate  the intuitive expectation  that incoherent
collisions  tend to drive the neutrino flavor field of the early universe  
toward homogeneous isotropic distributions. 
Our results suggest that  this heuristic picture may be  oversimplified.

At  small  scales,  corresponding to  Fourier modes $|\vec{k}| \gtrsim
100 \omega$, the growth of flavor instability can be  suppressed by
collisions, irrespective of the neutrino mass spectrum.  At large
length scales ($|\vec{k}| \lesssim 100 \omega$)    we find,  perhaps
surprisingly, that for inverted neutrino mass hierarchy
direction-changing collisions fail to  suppress flavor instabilities,
irrespective of the coupling strength.  In the extreme long-wavelength
case ($\vec{k}=0$) our numerical results are substantiated by an
analytic understanding of the problem (see
Section~\ref{linearstability}).  The key to understanding the puzzling
behavior is that for the simplified two-dimensional model, at large
scale ($|\vec{k}| \sim 0$) only $\theta$-independent modes grow
unstable for inverted hierarchy, while only $\theta$-dependent modes
grow unstable for normal hierarchy.

The appearance of inhomogeneity due to collective neutrino
oscillations in the early universe could have implications for the
entropy evolution in the neutrino sector, which in turn could affect  
Big Bang nucleosynthesis.  
Moreover,  the  non-trivial dependence of the instability  
(and its collisional suppression) on the neutrino mass hierarchy 
is  very intriguing and deserves attention in future  studies involving more 
realistic models.  

One obvious step towards a more realistic setup  would involve  
generalizing the collision kernels in order to account for 
energy-changing collisions in our formalism.  This would allow one to 
model neutrino interactions with electrons /  positrons in the early universe, 
which is beyond the scope of this Letter. 
However, even in absence of a complete calculation,  
since the collisional suppression of neutrino flavor oscillations 
 gets  contributions from direction change as well as  energy change, 
we speculate that the total effect of collisions on  the suppression of neutrino  inhomogeneities 
should  still be dependent on the neutrino mass hierarchy.

\section*{Acknowledgments}
The authors would like to thank George Fuller for offering valuable insights.
The authors would also like to thank Sajad Abbar, Daniel Blaschke,
Joe Carlson, Huaiyu Duan, Evan Grohs, Luke Johns, Lei Ma and Joshua Martin 
for numerous discussions.
The authors acknowledge support by the LDRD program at Los Alamos National Laboratory.

\section*{References}
\bibliographystyle{elsarticle-num}
\bibliography{2dgas}

\begin{thebibliography}{10}
\expandafter\ifx\csname url\endcsname\relax
  \def\url#1{\texttt{#1}}\fi
\expandafter\ifx\csname urlprefix\endcsname\relax\def\urlprefix{URL }\fi
\expandafter\ifx\csname href\endcsname\relax
  \def\href#1#2{#2} \def\path#1{#1}\fi

\bibitem{Duan:2014gfa}
H.~Duan, S.~Shalgar, {Flavor instabilities in the neutrino line model}, Phys.
  Lett. B747 (2015) 139--143.
\newblock \href {http://arxiv.org/abs/1412.7097} {\path{arXiv:1412.7097}},
  \href {http://dx.doi.org/10.1016/j.physletb.2015.05.057}
  {\path{doi:10.1016/j.physletb.2015.05.057}}.

\bibitem{Abbar:2015mca}
S.~Abbar, H.~Duan, S.~Shalgar, {Flavor instabilities in the multiangle neutrino
  line model}, Phys. Rev. D92~(6) (2015) 065019.
\newblock \href {http://arxiv.org/abs/1507.08992} {\path{arXiv:1507.08992}},
  \href {http://dx.doi.org/10.1103/PhysRevD.92.065019}
  {\path{doi:10.1103/PhysRevD.92.065019}}.

\bibitem{Mirizzi:2015fva}
A.~Mirizzi, G.~Mangano, N.~Saviano, {Self-induced flavor instabilities of a
  dense neutrino stream in a two-dimensional model}, Phys. Rev. D92~(2) (2015)
  021702.
\newblock \href {http://arxiv.org/abs/1503.03485} {\path{arXiv:1503.03485}},
  \href {http://dx.doi.org/10.1103/PhysRevD.92.021702}
  {\path{doi:10.1103/PhysRevD.92.021702}}.

\bibitem{Chakraborty:2015tfa}
S.~Chakraborty, R.~S. Hansen, I.~Izaguirre, G.~Raffelt, {Self-induced flavor
  conversion of supernova neutrinos on small scales}, JCAP 1601~(01) (2016)
  028.
\newblock \href {http://arxiv.org/abs/1507.07569} {\path{arXiv:1507.07569}},
  \href {http://dx.doi.org/10.1088/1475-7516/2016/01/028}
  {\path{doi:10.1088/1475-7516/2016/01/028}}.

\bibitem{Dolgov:1997mb}
A.~D. Dolgov, S.~H. Hansen, D.~V. Semikoz, {Nonequilibrium corrections to the
  spectra of massless neutrinos in the early universe}, Nucl. Phys. B503 (1997)
  426--444.
\newblock \href {http://arxiv.org/abs/hep-ph/9703315}
  {\path{arXiv:hep-ph/9703315}}, \href
  {http://dx.doi.org/10.1016/S0550-3213(97)00479-3}
  {\path{doi:10.1016/S0550-3213(97)00479-3}}.

\bibitem{Dolgov:2002ab}
A.~D. Dolgov, S.~H. Hansen, S.~Pastor, S.~T. Petcov, G.~G. Raffelt, D.~V.
  Semikoz, {Cosmological bounds on neutrino degeneracy improved by flavor
  oscillations}, Nucl. Phys. B632 (2002) 363--382.
\newblock \href {http://arxiv.org/abs/hep-ph/0201287}
  {\path{arXiv:hep-ph/0201287}}, \href
  {http://dx.doi.org/10.1016/S0550-3213(02)00274-2}
  {\path{doi:10.1016/S0550-3213(02)00274-2}}.

\bibitem{Mangano:2005cc}
G.~Mangano, G.~Miele, S.~Pastor, T.~Pinto, O.~Pisanti, P.~D. Serpico, {Relic
  neutrino decoupling including flavor oscillations}, Nucl. Phys. B729 (2005)
  221--234.
\newblock \href {http://arxiv.org/abs/hep-ph/0506164}
  {\path{arXiv:hep-ph/0506164}}, \href
  {http://dx.doi.org/10.1016/j.nuclphysb.2005.09.041}
  {\path{doi:10.1016/j.nuclphysb.2005.09.041}}.

\bibitem{Grohs:2015eua}
E.~Grohs, G.~M. Fuller, C.~T. Kishimoto, M.~W. Paris, {Probing neutrino physics
  with a self-consistent treatment of the weak decoupling, nucleosynthesis, and
  photon decoupling epochs}, JCAP 1505~(05) (2015) 017.
\newblock \href {http://arxiv.org/abs/1502.02718} {\path{arXiv:1502.02718}},
  \href {http://dx.doi.org/10.1088/1475-7516/2015/05/017}
  {\path{doi:10.1088/1475-7516/2015/05/017}}.

\bibitem{Grohs:2015tfy}
E.~Grohs, G.~M. Fuller, C.~T. Kishimoto, M.~W. Paris, A.~Vlasenko, {Neutrino
  energy transport in weak decoupling and big bang nucleosynthesis}, Phys. Rev.
  D93~(8) (2016) 083522.
\newblock \href {http://arxiv.org/abs/1512.02205} {\path{arXiv:1512.02205}},
  \href {http://dx.doi.org/10.1103/PhysRevD.93.083522}
  {\path{doi:10.1103/PhysRevD.93.083522}}.

\bibitem{Wolfenstein:1977ue}
L.~Wolfenstein, {Neutrino Oscillations in Matter}, Phys. Rev. D17 (1978)
  2369--2374.
\newblock \href {http://dx.doi.org/10.1103/PhysRevD.17.2369}
  {\path{doi:10.1103/PhysRevD.17.2369}}.

\bibitem{Mikheev:1986gs}
S.~P. Mikheev, A.~{\relax Yu}. Smirnov, {Resonance Amplification of
  Oscillations in Matter and Spectroscopy of Solar Neutrinos}, Sov. J. Nucl.
  Phys. 42 (1985) 913--917, [Yad. Fiz.42,1441(1985)].

\bibitem{Pantaleone:1994ns}
J.~T. Pantaleone, {Neutrino flavor evolution near a supernova's core}, Phys.
  Lett. B342 (1995) 250--256.
\newblock \href {http://arxiv.org/abs/astro-ph/9405008}
  {\path{arXiv:astro-ph/9405008}}, \href
  {http://dx.doi.org/10.1016/0370-2693(94)01369-N}
  {\path{doi:10.1016/0370-2693(94)01369-N}}.

\bibitem{Duan:2005cp}
H.~Duan, G.~M. Fuller, Y.-Z. Qian, {Collective neutrino flavor transformation
  in supernovae}, Phys. Rev. D74 (2006) 123004.
\newblock \href {http://arxiv.org/abs/astro-ph/0511275}
  {\path{arXiv:astro-ph/0511275}}, \href
  {http://dx.doi.org/10.1103/PhysRevD.74.123004}
  {\path{doi:10.1103/PhysRevD.74.123004}}.

\bibitem{Duan:2006an}
H.~Duan, G.~M. Fuller, J.~Carlson, Y.-Z. Qian, {Simulation of Coherent
  Non-Linear Neutrino Flavor Transformation in the Supernova Environment. 1.
  Correlated Neutrino Trajectories}, Phys. Rev. D74 (2006) 105014.
\newblock \href {http://arxiv.org/abs/astro-ph/0606616}
  {\path{arXiv:astro-ph/0606616}}, \href
  {http://dx.doi.org/10.1103/PhysRevD.74.105014}
  {\path{doi:10.1103/PhysRevD.74.105014}}.

\bibitem{Duan:2006jv}
H.~Duan, G.~M. Fuller, J.~Carlson, Y.-Z. Qian, {Coherent Development of
  Neutrino Flavor in the Supernova Environment}, Phys. Rev. Lett. 97 (2006)
  241101.
\newblock \href {http://arxiv.org/abs/astro-ph/0608050}
  {\path{arXiv:astro-ph/0608050}}, \href
  {http://dx.doi.org/10.1103/PhysRevLett.97.241101}
  {\path{doi:10.1103/PhysRevLett.97.241101}}.

\bibitem{Duan:2007mv}
H.~Duan, G.~M. Fuller, J.~Carlson, Y.-Z. Qian, {Analysis of Collective Neutrino
  Flavor Transformation in Supernovae}, Phys. Rev. D75 (2007) 125005.
\newblock \href {http://arxiv.org/abs/astro-ph/0703776}
  {\path{arXiv:astro-ph/0703776}}, \href
  {http://dx.doi.org/10.1103/PhysRevD.75.125005}
  {\path{doi:10.1103/PhysRevD.75.125005}}.

\bibitem{Duan:2008za}
H.~Duan, G.~M. Fuller, Y.-Z. Qian, {Stepwise spectral swapping with three
  neutrino flavors}, Phys. Rev. D77 (2008) 085016.
\newblock \href {http://arxiv.org/abs/0801.1363} {\path{arXiv:0801.1363}},
  \href {http://dx.doi.org/10.1103/PhysRevD.77.085016}
  {\path{doi:10.1103/PhysRevD.77.085016}}.

\bibitem{Raffelt:2007xt}
G.~G. Raffelt, A.~{\relax Yu}. Smirnov, {Adiabaticity and spectral splits in
  collective neutrino transformations}, Phys. Rev. D76 (2007) 125008.
\newblock \href {http://arxiv.org/abs/0709.4641} {\path{arXiv:0709.4641}},
  \href {http://dx.doi.org/10.1103/PhysRevD.76.125008}
  {\path{doi:10.1103/PhysRevD.76.125008}}.

\bibitem{EstebanPretel:2007bz}
A.~Esteban-Pretel, S.~Pastor, R.~Tomas, G.~G. Raffelt, G.~Sigl, {Multi-angle
  effects in collective supernova neutrino oscillations}, J. Phys. Conf. Ser.
  120 (2008) 052021.
\newblock \href {http://arxiv.org/abs/0712.2176} {\path{arXiv:0712.2176}},
  \href {http://dx.doi.org/10.1088/1742-6596/120/5/052021}
  {\path{doi:10.1088/1742-6596/120/5/052021}}.

\bibitem{EstebanPretel:2008ni}
A.~Esteban-Pretel, A.~Mirizzi, S.~Pastor, R.~Tomas, G.~G. Raffelt, P.~D.
  Serpico, G.~Sigl, {Role of dense matter in collective supernova neutrino
  transformations}, Phys. Rev. D78 (2008) 085012.
\newblock \href {http://arxiv.org/abs/0807.0659} {\path{arXiv:0807.0659}},
  \href {http://dx.doi.org/10.1103/PhysRevD.78.085012}
  {\path{doi:10.1103/PhysRevD.78.085012}}.

\bibitem{Raffelt:2008hr}
G.~G. Raffelt, {Self-induced parametric resonance in collective neutrino
  oscillations}, Phys. Rev. D78 (2008) 125015.
\newblock \href {http://arxiv.org/abs/0810.1407} {\path{arXiv:0810.1407}},
  \href {http://dx.doi.org/10.1103/PhysRevD.78.125015}
  {\path{doi:10.1103/PhysRevD.78.125015}}.

\bibitem{Dasgupta:2010ae}
B.~Dasgupta, G.~G. Raffelt, I.~Tamborra, {Triggering collective oscillations by
  three-flavor effects}, Phys. Rev. D81 (2010) 073004.
\newblock \href {http://arxiv.org/abs/1001.5396} {\path{arXiv:1001.5396}},
  \href {http://dx.doi.org/10.1103/PhysRevD.81.073004}
  {\path{doi:10.1103/PhysRevD.81.073004}}.

\bibitem{Malkus:2014iqa}
A.~Malkus, A.~Friedland, G.~C. McLaughlin, {Matter-Neutrino Resonance Above
  Merging Compact Objects}\href {http://arxiv.org/abs/1403.5797}
  {\path{arXiv:1403.5797}}.

\bibitem{Zhu:2016mwa}
Y.-L. Zhu, A.~Perego, G.~C. McLaughlin, {Matter Neutrino Resonance Transitions
  above a Neutron Star Merger Remnant}, Phys. Rev. D94~(10) (2016) 105006.
\newblock \href {http://arxiv.org/abs/1607.04671} {\path{arXiv:1607.04671}},
  \href {http://dx.doi.org/10.1103/PhysRevD.94.105006}
  {\path{doi:10.1103/PhysRevD.94.105006}}.

\bibitem{Wu:2015fga}
M.-R. Wu, H.~Duan, Y.-Z. Qian, {Physics of neutrino flavor transformation
  through matter–neutrino resonances}, Phys. Lett. B752 (2016) 89--94.
\newblock \href {http://arxiv.org/abs/1509.08975} {\path{arXiv:1509.08975}},
  \href {http://dx.doi.org/10.1016/j.physletb.2015.11.027}
  {\path{doi:10.1016/j.physletb.2015.11.027}}.

\bibitem{Raffelt:2013rqa}
G.~Raffelt, S.~Sarikas, D.~de~Sousa~Seixas, {Axial Symmetry Breaking in
  Self-Induced Flavor Conversionof Supernova Neutrino Fluxes}, Phys. Rev. Lett.
  111~(9) (2013) 091101, [Erratum: Phys. Rev. Lett.113,no.23,239903(2014)].
\newblock \href {http://arxiv.org/abs/1305.7140} {\path{arXiv:1305.7140}},
  \href {http://dx.doi.org/10.1103/PhysRevLett.113.239903,
  10.1103/PhysRevLett.111.091101} {\path{doi:10.1103/PhysRevLett.113.239903,
  10.1103/PhysRevLett.111.091101}}.

\bibitem{Mirizzi:2013wda}
S.~Chakraborty, A.~Mirizzi, {Multi-azimuthal-angle instability for different
  supernova neutrino fluxes}, Phys. Rev. D90~(3) (2014) 033004.
\newblock \href {http://arxiv.org/abs/1308.5255} {\path{arXiv:1308.5255}},
  \href {http://dx.doi.org/10.1103/PhysRevD.90.033004}
  {\path{doi:10.1103/PhysRevD.90.033004}}.

\bibitem{Mangano:2001iu}
G.~Mangano, G.~Miele, S.~Pastor, M.~Peloso, {A Precision calculation of the
  effective number of cosmological neutrinos}, Phys. Lett. B534 (2002) 8--16.
\newblock \href {http://arxiv.org/abs/astro-ph/0111408}
  {\path{arXiv:astro-ph/0111408}}, \href
  {http://dx.doi.org/10.1016/S0370-2693(02)01622-2}
  {\path{doi:10.1016/S0370-2693(02)01622-2}}.

\bibitem{deSalas:2016ztq}
P.~F. de~Salas, S.~Pastor, {Relic neutrino decoupling with flavour oscillations
  revisited}, JCAP 1607~(07) (2016) 051.
\newblock \href {http://arxiv.org/abs/1606.06986} {\path{arXiv:1606.06986}}.

\bibitem{Harvey:1981cu}
J.~A. Harvey, E.~W. Kolb, {Grand Unified Theories and the Lepton Number of the
  Universe}, Phys. Rev. D24 (1981) 2090.
\newblock \href {http://dx.doi.org/10.1103/PhysRevD.24.2090}
  {\path{doi:10.1103/PhysRevD.24.2090}}.

\bibitem{Foot:1995qk}
R.~Foot, M.~J. Thomson, R.~R. Volkas, {Large neutrino asymmetries from neutrino
  oscillations}, Phys. Rev. D53 (1996) R5349--R5353.
\newblock \href {http://arxiv.org/abs/hep-ph/9509327}
  {\path{arXiv:hep-ph/9509327}}, \href
  {http://dx.doi.org/10.1103/PhysRevD.53.R5349}
  {\path{doi:10.1103/PhysRevD.53.R5349}}.

\bibitem{Shi:1996ic}
X.-D. Shi, {Chaotic amplification of neutrino chemical potentials by neutrino
  oscillations in big bang nucleosynthesis}, Phys. Rev. D54 (1996) 2753--2760.
\newblock \href {http://arxiv.org/abs/astro-ph/9602135}
  {\path{arXiv:astro-ph/9602135}}, \href
  {http://dx.doi.org/10.1103/PhysRevD.54.2753}
  {\path{doi:10.1103/PhysRevD.54.2753}}.

\bibitem{Casas:1997gx}
A.~Casas, W.~Y. Cheng, G.~Gelmini, {Generation of large lepton asymmetries},
  Nucl. Phys. B538 (1999) 297--308.
\newblock \href {http://arxiv.org/abs/hep-ph/9709289}
  {\path{arXiv:hep-ph/9709289}}, \href
  {http://dx.doi.org/10.1016/S0550-3213(98)00606-3}
  {\path{doi:10.1016/S0550-3213(98)00606-3}}.

\bibitem{MarchRussell:1999ig}
J.~March-Russell, H.~Murayama, A.~Riotto, {The Small observed baryon asymmetry
  from a large lepton asymmetry}, JHEP 11 (1999) 015.
\newblock \href {http://arxiv.org/abs/hep-ph/9908396}
  {\path{arXiv:hep-ph/9908396}}, \href
  {http://dx.doi.org/10.1088/1126-6708/1999/11/015}
  {\path{doi:10.1088/1126-6708/1999/11/015}}.

\bibitem{Kawasaki:2002hq}
M.~Kawasaki, F.~Takahashi, M.~Yamaguchi, {Large lepton asymmetry from Q balls},
  Phys. Rev. D66 (2002) 043516.
\newblock \href {http://arxiv.org/abs/hep-ph/0205101}
  {\path{arXiv:hep-ph/0205101}}, \href
  {http://dx.doi.org/10.1103/PhysRevD.66.043516}
  {\path{doi:10.1103/PhysRevD.66.043516}}.

\bibitem{Yamaguchi:2002vw}
M.~Yamaguchi, {Generation of cosmological large lepton asymmetry from a rolling
  scalar field}, Phys. Rev. D68 (2003) 063507.
\newblock \href {http://arxiv.org/abs/hep-ph/0211163}
  {\path{arXiv:hep-ph/0211163}}, \href
  {http://dx.doi.org/10.1103/PhysRevD.68.063507}
  {\path{doi:10.1103/PhysRevD.68.063507}}.

\bibitem{Shaposhnikov:2008pf}
M.~Shaposhnikov, {The nuMSM, leptonic asymmetries, and properties of singlet
  fermions}, JHEP 08 (2008) 008.
\newblock \href {http://arxiv.org/abs/0804.4542} {\path{arXiv:0804.4542}},
  \href {http://dx.doi.org/10.1088/1126-6708/2008/08/008}
  {\path{doi:10.1088/1126-6708/2008/08/008}}.

\bibitem{Gu:2010dg}
P.-H. Gu, {Large Lepton Asymmetry for Small Baryon Asymmetry and Warm Dark
  Matter}, Phys. Rev. D82 (2010) 093009.
\newblock \href {http://arxiv.org/abs/1005.1632} {\path{arXiv:1005.1632}},
  \href {http://dx.doi.org/10.1103/PhysRevD.82.093009}
  {\path{doi:10.1103/PhysRevD.82.093009}}.

\bibitem{Johns:2016enc}
L.~Johns, M.~Mina, V.~Cirigliano, M.~W. Paris, G.~M. Fuller, {Neutrino flavor
  transformation in the lepton-asymmetric universe}, Phys. Rev. D94~(8) (2016)
  083505.
\newblock \href {http://arxiv.org/abs/1608.01336} {\path{arXiv:1608.01336}},
  \href {http://dx.doi.org/10.1103/PhysRevD.94.083505}
  {\path{doi:10.1103/PhysRevD.94.083505}}.

\bibitem{Grohs:2016vef}
E.~Grohs, G.~M. Fuller, {The surprising influence of late charged current weak
  interactions on Big Bang Nucleosynthesis}, Nucl. Phys. B911 (2016) 955--973.
\newblock \href {http://arxiv.org/abs/1607.02797} {\path{arXiv:1607.02797}},
  \href {http://dx.doi.org/10.1016/j.nuclphysb.2016.08.034}
  {\path{doi:10.1016/j.nuclphysb.2016.08.034}}.

\bibitem{Grohs:2016cuu}
E.~Grohs, G.~M. Fuller, C.~T. Kishimoto, M.~W. Paris, {Lepton asymmetry,
  neutrino spectral distortions, and big bang nucleosynthesis}, Phys. Rev.
  D95~(6) (2017) 063503.
\newblock \href {http://arxiv.org/abs/1612.01986} {\path{arXiv:1612.01986}},
  \href {http://dx.doi.org/10.1103/PhysRevD.95.063503}
  {\path{doi:10.1103/PhysRevD.95.063503}}.

\bibitem{Cirigliano:2014aoa}
V.~Cirigliano, G.~M. Fuller, A.~Vlasenko, {A New Spin on Neutrino Quantum
  Kinetics}, Phys. Lett. B747 (2015) 27--35.
\newblock \href {http://arxiv.org/abs/1406.5558} {\path{arXiv:1406.5558}},
  \href {http://dx.doi.org/10.1016/j.physletb.2015.04.066}
  {\path{doi:10.1016/j.physletb.2015.04.066}}.

\bibitem{Vlasenko:2013fja}
A.~Vlasenko, G.~M. Fuller, V.~Cirigliano, {Neutrino Quantum Kinetics}, Phys.
  Rev. D89~(10) (2014) 105004.
\newblock \href {http://arxiv.org/abs/1309.2628} {\path{arXiv:1309.2628}},
  \href {http://dx.doi.org/10.1103/PhysRevD.89.105004}
  {\path{doi:10.1103/PhysRevD.89.105004}}.

\bibitem{Blaschke:2016xxt}
D.~N. Blaschke, V.~Cirigliano, {Neutrino Quantum Kinetic Equations: The
  Collision Term}, Phys. Rev. D94~(3) (2016) 033009.
\newblock \href {http://arxiv.org/abs/1605.09383} {\path{arXiv:1605.09383}},
  \href {http://dx.doi.org/10.1103/PhysRevD.94.033009}
  {\path{doi:10.1103/PhysRevD.94.033009}}.

\bibitem{Sigl:1992fn}
G.~Sigl, G.~Raffelt, {General kinetic description of relativistic mixed
  neutrinos}, Nucl. Phys. B406 (1993) 423--451.
\newblock \href {http://dx.doi.org/10.1016/0550-3213(93)90175-O}
  {\path{doi:10.1016/0550-3213(93)90175-O}}.

\bibitem{Raffelt:1992uj}
G.~Raffelt, G.~Sigl, L.~Stodolsky, {NonAbelian Boltzmann equation for mixing
  and decoherence}, Phys. Rev. Lett. 70 (1993) 2363--2366, [Erratum: Phys. Rev.
  Lett.98,069902(2007)].
\newblock \href {http://arxiv.org/abs/hep-ph/9209276}
  {\path{arXiv:hep-ph/9209276}}, \href
  {http://dx.doi.org/10.1103/PhysRevLett.70.2363,
  10.1103/PhysRevLett.98.069902} {\path{doi:10.1103/PhysRevLett.70.2363,
  10.1103/PhysRevLett.98.069902}}.

\bibitem{McKellar:1992ja}
B.~H.~J. McKellar, M.~J. Thomson, {Oscillating doublet neutrinos in the early
  universe}, Phys. Rev. D49 (1994) 2710--2728.
\newblock \href {http://dx.doi.org/10.1103/PhysRevD.49.2710}
  {\path{doi:10.1103/PhysRevD.49.2710}}.

\bibitem{Enqvist:1990ad}
K.~Enqvist, K.~Kainulainen, J.~Maalampi, {Refraction and Oscillations of
  Neutrinos in the Early Universe}, Nucl. Phys. B349 (1991) 754--790.
\newblock \href {http://dx.doi.org/10.1016/0550-3213(91)90397-G}
  {\path{doi:10.1016/0550-3213(91)90397-G}}.

\bibitem{Strack:2005ux}
P.~Strack, A.~Burrows, {Generalized Boltzmann formalism for oscillating
  neutrinos}, Phys. Rev. D71 (2005) 093004.
\newblock \href {http://arxiv.org/abs/hep-ph/0504035}
  {\path{arXiv:hep-ph/0504035}}, \href
  {http://dx.doi.org/10.1103/PhysRevD.71.093004}
  {\path{doi:10.1103/PhysRevD.71.093004}}.

\bibitem{Volpe:2013jgr}
C.~Volpe, D.~Väänänen, C.~Espinoza, {Extended evolution equations for
  neutrino propagation in astrophysical and cosmological environments}, Phys.
  Rev. D87~(11) (2013) 113010.
\newblock \href {http://arxiv.org/abs/1302.2374} {\path{arXiv:1302.2374}},
  \href {http://dx.doi.org/10.1103/PhysRevD.87.113010}
  {\path{doi:10.1103/PhysRevD.87.113010}}.

\bibitem{Volpe:2015rla}
C.~Volpe, {Neutrino Quantum Kinetic Equations}, Int. J. Mod. Phys. E24~(09)
  (2015) 1541009.
\newblock \href {http://arxiv.org/abs/1506.06222} {\path{arXiv:1506.06222}},
  \href {http://dx.doi.org/10.1142/S0218301315410098}
  {\path{doi:10.1142/S0218301315410098}}.

\bibitem{Zhang:2013lka}
Y.~Zhang, A.~Burrows, {Transport Equations for Oscillating Neutrinos}, Phys.
  Rev. D88~(10) (2013) 105009.
\newblock \href {http://arxiv.org/abs/1310.2164} {\path{arXiv:1310.2164}},
  \href {http://dx.doi.org/10.1103/PhysRevD.88.105009}
  {\path{doi:10.1103/PhysRevD.88.105009}}.

\bibitem{Serreau:2014cfa}
J.~Serreau, C.~Volpe, {Neutrino-antineutrino correlations in dense anisotropic
  media}, Phys. Rev. D90~(12) (2014) 125040.
\newblock \href {http://arxiv.org/abs/1409.3591} {\path{arXiv:1409.3591}},
  \href {http://dx.doi.org/10.1103/PhysRevD.90.125040}
  {\path{doi:10.1103/PhysRevD.90.125040}}.

\bibitem{Mangano:2011ip}
G.~Mangano, G.~Miele, S.~Pastor, O.~Pisanti, S.~Sarikas, {Updated BBN bounds on
  the cosmological lepton asymmetry for non-zero $\theta_{13}$}, Phys. Lett.
  B708 (2012) 1--5.
\newblock \href {http://arxiv.org/abs/1110.4335} {\path{arXiv:1110.4335}},
  \href {http://dx.doi.org/10.1016/j.physletb.2012.01.015}
  {\path{doi:10.1016/j.physletb.2012.01.015}}.

\bibitem{Castorina:2012md}
E.~Castorina, U.~Franca, M.~Lattanzi, J.~Lesgourgues, G.~Mangano,
  A.~Melchiorri, S.~Pastor, {Cosmological lepton asymmetry with a nonzero
  mixing angle $\theta_{13}$}, Phys. Rev. D86 (2012) 023517.
\newblock \href {http://arxiv.org/abs/1204.2510} {\path{arXiv:1204.2510}},
  \href {http://dx.doi.org/10.1103/PhysRevD.86.023517}
  {\path{doi:10.1103/PhysRevD.86.023517}}.

\bibitem{Pastor:2008ti}
S.~Pastor, T.~Pinto, G.~G. Raffelt, {Relic density of neutrinos with primordial
  asymmetries}, Phys. Rev. Lett. 102 (2009) 241302.
\newblock \href {http://arxiv.org/abs/0808.3137} {\path{arXiv:0808.3137}},
  \href {http://dx.doi.org/10.1103/PhysRevLett.102.241302}
  {\path{doi:10.1103/PhysRevLett.102.241302}}.

\bibitem{Mangano:2010ei}
G.~Mangano, G.~Miele, S.~Pastor, O.~Pisanti, S.~Sarikas, {Constraining the
  cosmic radiation density due to lepton number with Big Bang Nucleosynthesis},
  JCAP 1103 (2011) 035.
\newblock \href {http://arxiv.org/abs/1011.0916} {\path{arXiv:1011.0916}},
  \href {http://dx.doi.org/10.1088/1475-7516/2011/03/035}
  {\path{doi:10.1088/1475-7516/2011/03/035}}.

\bibitem{Banerjee:2011fj}
A.~Banerjee, A.~Dighe, G.~Raffelt, {Linearized flavor-stability analysis of
  dense neutrino streams}, Phys. Rev. D84 (2011) 053013.
\newblock \href {http://arxiv.org/abs/1107.2308} {\path{arXiv:1107.2308}},
  \href {http://dx.doi.org/10.1103/PhysRevD.84.053013}
  {\path{doi:10.1103/PhysRevD.84.053013}}.

\bibitem{Gough:2009:GSL:1538674}
B.~Gough, GNU Scientific Library Reference Manual - Third Edition, 3rd Edition,
  Network Theory Ltd., 2009.

\end{thebibliography}

\end{document}